\documentclass[aps,prd,twocolumn,amsmath,superscriptaddress,amssymb,showpacs,floatfix,nofootinbib,longbibliography]{revtex4-1}
\usepackage{graphicx}
\usepackage{dcolumn}
\usepackage{xcolor}
\usepackage{bm}
\usepackage{calligra}
\usepackage[T1]{fontenc}
\usepackage{egothic}
\usepackage[T1]{fontenc}
\newfont{\rsfsten}{rsfs10 scaled 1200}
\newfont{\rsfsseven}{rsfs10 scaled 1200}
\newfont{\rsfsfive}{rsfs10 scaled 1200}
\usepackage{epsfig}
\usepackage{units}
\usepackage[utf8]{inputenc}
\usepackage{hyperref}
\usepackage{tabularx}
\newcommand{\be}{\begin{equation}}
\newcommand{\ee}{\end{equation}}
\newcommand{\bea}{\begin{eqnarray}}
\newcommand{\eea}{\end{eqnarray}}

\hypersetup{
colorlinks=true,
citecolor=darkblue,
linkcolor=reddish,
urlcolor=darkblue,
pdfauthor={},
pdftitle={},
pdfsubject={}
}
\newcommand{\mnras}{MNRAS}
\newcommand{\aap}{Astronomy and Astrophysics}

\definecolor{blue2}{cmyk}{1, 0.1, 0.1, 0.1}

\usepackage{colortbl}
\definecolor{lightgreen}{cmyk}{0.2, 0, 0.2, 0.2}
\definecolor{lightgray2}{cmyk}{0.1,0.1,0,0.1}
\definecolor{Red2}{RGB}{214, 39, 40}
\definecolor{Blue2}{RGB} {31, 119, 180}
\definecolor{Orange2}{RGB}{255, 127, 14}
\definecolor{Green2}{RGB}{44, 160, 44}
\definecolor{greyish2}{rgb}{.96,.96,.96}

\definecolor{Red}{RGB}{214, 39, 40}
\definecolor{Blue}{RGB} {31, 119, 180}
\definecolor{Orange}{RGB}{255, 153, 51}
\definecolor{Purple}{RGB}{178, 102, 255}
\definecolor{Green}{RGB}{44, 160, 44}
\definecolor{regal}{RGB}{90,0,120}
\definecolor{darkblue}{rgb}{0.15,0.35,0.55}
\definecolor{reddish}{rgb}{0.65, 0.2, 0.2}
\definecolor{darkgreen}{RGB}{50,150,0}
\definecolor{greyish}{rgb}{.90,.90,.90}
\definecolor{greyish2}{rgb}{.96,.96,.96}
\definecolor{greyish3}{rgb}{.37,.37,.37}
\definecolor{darkblue2}{rgb}{0.3,0.4,0.9}
\definecolor{Blue3}{RGB}{31, 119, 180}

\definecolor{pyBlue}{RGB}{31, 119, 180}
\definecolor{pyRed}{RGB}{214, 39, 40}
\definecolor{pyGreen}{RGB}{44, 160, 44}
\definecolor{pyBlue2}{RGB}{0, 111, 237}
\definecolor{pyRed2}{RGB}{224, 52, 36}
\definecolor{Mathematica1}{rgb}{0.368417, 0.506779, 0.709798}
\definecolor{Mathematica2}{rgb}{0.880722, 0.611041, 0.142051}
\usepackage{ragged2e}

\usepackage{comment}

\newcommand{\RNum}[1]{\uppercase\expandafter{\romannumeral #1\relax}}

\begin{document}

\title{Conservative limits on primordial black holes from the LIGO-Virgo-KAGRA observations}

\author{Mehdi El Bouhaddouti}
\email{melbouhaddouti@oakland.edu, ORCID: orcid.org/0009-0001-1299-0879}
\affiliation{Department of physics, Oakland University, Rochester Michigan, 48309,U.S.A}

\author{Muhsin Aljaf}
\email{muhsinaljaf@oakland.edu, ORCID: orcid.org/0000-0002-1251-4928}
\affiliation{Department of physics, Oakland University, Rochester Michigan, 48309,U.S.A}

\author{Ilias Cholis}
\email{cholis@oakland.edu, ORCID: orcid.org/0000-0002-3805-6478}
\affiliation{Department of physics, Oakland University, Rochester Michigan, 48309,U.S.A}

\begin{abstract} 
Primordial black holes (PBH) may constitute a considerable fraction of dark matter. In this work we use the recent observations by the LIGO-Virgo-KAGRA (LVK) collaborations to set direct limits on stellar-mass range PBHs. We evaluate the merger rates of PBH binaries by accounting for the binaries formed by two-body captures inside dark matter halos and by studying the evolution of PBH binaries inside such halos through binary-single interactions. Those type of interactions contribute to what is a minimum of PBH merger rates at low redshifts detectable by LVK. Thus, they allow us to derive what is the most conservative upper limits on the presence of merging PBH binaries in the gravitational-wave observations. We study both the case where PBHs have a monochromatic mass-distribution and the case where that distribution is described by a log-normal function. To derive limits on PBHs, we simulate binaries with masses following the relevant mass-distributions and merger rates as a function of redshift, up to $z\sim 0.8$, relevant for the current LVK observations. In those simulations we also take into account that the LVK observatories measure the detected black hole binaries' masses with a finite resolution. In comparing to the LVK observations, we combine the simulated PBH binaries with the binaries following a power-law mass- and redshift-distribution. The latter binaries dominate the observed population of LVK detections. Our derived limits on the mass fraction of dark matter composed of PBHs is in the range of $10^{-4}$ to $2\times 10^{-2}$, depending on the exact assumptions relating to the PBH binaries properties. For reasonable assumptions on those PBH binaries' properties before their evolution inside dark matter halos, we get that fraction to be in the range of $10^{-3} - 10^{-2}$, for PBH masses of 5-80 $M_{\odot}$. Our work provide some of the most competitive limits in the mass range of 5-50 $M_{\odot}$.    
\end{abstract}

\maketitle

\section{Introduction} 
\label{sec:intro}
The nature of dark matter remains one of the most fundamental open questions in cosmology and particle physics. Candidates, in mass, span many orders of magnitude, ranging from $10^{-22}$ eV to $O(10^3)\, M_{\odot}$ \cite{Battaglieri:2017aum, Boddy:2022knd}. The highest-end of the dark matter candidates' mass range, is occupied by primordial black holes (PBHs), generated from the collapse of early universe curvature perturbations \cite{1974MNRAS.168..399C, 1974A&A....37..225M, 1975ApJ...201....1C}. 
With the current generation of ground-based gravitational-wave (GW) observatories we have the opportunity, for the first time to directly probe PBHs with masses $\gtrsim 1 M_{\odot}$, through the merger of PBH binaries leaving a signal at the LIGO-Virgo-KAGRA (LVK) interferometers \cite{Bird:2016dcv, Sasaki:2016jop, Clesse:2016vqa, Bertone:2018krk, Sasaki:2018dmp, Carr:2020xqk}. 

The most direct manner by which we can probe a signal of PBHs in the LVK observations is through the mass-distribution of the detected events containing black holes (BHs) and other compact objects \cite{Kovetz:2016kpi, Gow:2019pok, Kovetz:2017rvv}. These include studying either only the individual masses of the observed BHs \cite{Gupta:2019nwj, Golomb:2023vxm}, or the combined information of the binaries' masses \cite{Kovetz:2016kpi, Hall:2020daa, Chen:2024dxh}. In the latter case, one cares also on how correlated the values of the primary mass of the binary $M_{1}$ is to the secondary mass of the binary $M_{2}$; as that may provide a manner to separate the process by which these binaries have formed (see e.g. \cite{Kovetz:2016kpi, Dvorkin:2017kfg, Flitter:2020bky, Wang:2020aoh, Cholis:2021dzh, Golomb:2023vxm}). Alternative probes/signals of a PBH component in the GW observations have also been proposed in \cite{Cholis:2016xvo, Cholis:2016kqi, Raccanelli:2016cud, Mandic:2016lcn, Gerosa:2018wbw, Clesse:2016ajp, Nakama:2016gzw}. 

PBH binaries that merge may form in multiple pathways \cite{Khlopov_2010}. One path is through, GW direct captures, which happens when two unbounded PBHs come that close to emit enough GW energy to form a highly-eccentric binary that would merge soon after \cite{Bird:2016dcv}. 
A second pathway is PBH binaries may form from the early universe perturbations and evolve slowly through GW emission to merge in our time without interacting substantively with their environments \cite{Sasaki:2016jop, Ali-Haimoud:2017rtz} (see also Ref.~\cite{Raidal:2017mfl}).
Finally a third path to binaries that merge within the age of the universe is  via interactions occurring between existing PBH binaries and single PBHs within the dark matter halos \cite{Franciolini:2022ewd, Aljaf:2024fru, Aljaf:2025dta}. These dynamical effects inside dark matter halos can accelerate or delay the orbital evolution of binaries compared to the evolution purely through gravitational-wave emission (see Ref.~\cite{ Aljaf:2025dta}). 
At least $\simeq 90\%$ of all dark matter in the universe has been inside halos since redshift of 2. Thus, the calculations on the PBH merger rates of \cite{Franciolini:2022ewd, Aljaf:2024fru, Aljaf:2025dta}, do take into account the merger of binaries that in \cite{Sasaki:2016jop, Ali-Haimoud:2017rtz, Raidal:2017mfl} are assumed to evolve undisturbed simply through GW emission 
That is especially important for the limits derived based on the current LVK measurements that only probe redshifts of $\lesssim 1$. 

In this paper, we use the LVK observations from observing runs 1 (O1) to 3 (O3), to place limits on the abundance of PBHs. We describe those in terms of the fraction dark matter density in PBHs, $f_{\textrm{PBH}}$, where $\rho_{\textrm{PBH}} = f_{\textrm{PBH}} \cdot \rho_{\textrm{DM}}$. 
In Sec.~\ref{sec:rates}, we describe our calculations of the PBH merger rates under assumptions on their initial abundance, the fraction of PBH in binaries and their mass distribution. 
To derive our limits in Sec.~\ref{sec:BBHs_limits}, we rely on the method of representing the O1-O3 LVK observations presented in \cite{Bouhaddouti:2024ena}. We place limits on PBHs, first under the assumption that they have a monochromatic mass-distribution, with that mass being a free parameter. We find that the O1-O3 LVK limits set some of the most robust and tight limits in the mass range of 3 to $40 \, M_{\odot}$, with $1 \times 10^{-3} < f_{\textrm{PBH}} < 5 \times 10^{-3}$. We also study the assumption that PBHs follow a log-normal mass-distribution \cite{Kannike:2017bxn, Gow:2020cou}. Our limits for the log-normal distribution are also the most competitive among the various probes; between 5 and $50 \, M_{\odot}$, we get $5 \times 10^{-4} < f_{\textrm{PBH}} < 2 \times 10^{-3}$. Alternative mass distributions for the PBHs have been proposed as in \cite{Niemeyer:1997mt, Yokoyama:1998xd, Carr:2016drx, DeLuca:2020ioi, Deng:2021ezy} give similar limits. 
We conclude in Sec.~\ref{sec:conclusions} and discuss future improvements in probing for a PBH signal in the black hole mass distribution with upcoming GW observations.

\section{The merger rates of primordial black holes}
In this section we calculate PBH binary mergers from three distinct channels: direct captures  between PBHs inside halos, mergers from binary-single interactions between in halos and unperturbed binaries residing outside halos. 
In this work we consider that PBHs exist both as single objects and in binaries with other PBHs. The fraction of PBHs in binaries is $0< f_{\mathrm{PBH \, binaries}} < 1$. 
\label{sec:rates}

\subsection{Monochromatic primordial black holes}
\label{subsec:rates_monochromatic}

For a given dark matter halo or virial radius $R_{\textrm{vir}}$, the direct PBH capture rate  is \cite{Bird:2016dcv,Aljaf:2024fru},
\begin{equation}\label{captures_eq}
\Gamma_{\mathrm{capture}} = 2\pi \int_{0}^{R_{\mathrm{vir}}} r^2 \langle \sigma \cdot v_{\mathrm{PBH}} \rangle \frac{[f_{\textrm{PBH}} \cdot \rho_{\mathrm{NFW}}(r)]^2}{m_{\textrm{PBH}}^2} \, dr,
\end{equation}
where $\sigma$ is the cross-section for direct captures \cite{1989ApJ...343..725Q, Mouri:2002mc}, $m_{\textrm{PBH}}$ is the PBH mass, $v_{\mathrm{PBH}}$ is relative velocity between them and 
$\rho_{\mathrm{NFW}}$ is density profile of the dark matter halo. We assume that all dark matter halos follow the Navarro, Frenk and White (NFW) density profile \cite{1997ApJ...490..493N}. 
Given the cosmologically short interval between the creation of these binaries via captures and the binaries' mergers, this leads to the capture merger rate per halo, $R_{\mathrm{captures}}^{\mathrm{halo}}(M,z)$, given by,
\begin{eqnarray}
R_{\mathrm{captures}}^{\mathrm{halo}}(M,z) &=& \frac{2 \pi }{3}\left(\frac{85 \pi}{6 \sqrt{2}}\right)^{2 / 7} f_{\mathrm{PBH}}^2 \\
&\times& \frac{G^2 \cdot \left[ M_{\mathrm{vir}}(z) \right]^2 \cdot  D_v(z) \cdot f(C(z))}{\left[ R_s(z) \right]^3 \cdot c  \cdot \left[g(C(z))\right]^2 }. \nonumber
\end{eqnarray}
$M_{\mathrm{vir}}(z)$, is the virial mass of a given dark matter halo at redshift $z$ with a scale radius $R_s(z)$. $D_v(z)$ is related to the velocity distribution of PBHs inside halos. $C(z)$ is the redshift-dependent concentration parameter of dark matter halos, and $g(C(z))$ and $f(C(z))$ are relevant functions (see Appendix \ref{app:halo_properries} and Ref. \cite{Aljaf:2024fru} for more details).
The capture rate across halos is then calculated as,
\begin{equation}
R_{\textrm{captures}}(M, z)=\int_{M_{\textrm{min}}}^{M_{\textrm{max}}} R_{\text{captures}}^{\text{halo}}(M, z) \frac{d n}{d M} d M,
\label{eq:HMF}
\end{equation}
where $dn/dM$ is the halo mass function. We use the Press-Schechter halo mass function \cite{1974ApJ...187..425P} and consider dark matter halos in the mass range of $10^4 - 10^{15} \, M_\odot$. A detailed calculation for the case where the PBHs follow a continuous mass-distribution is provided in  Ref.~\cite{Aljaf:2024fru}. We note that for direct captures, the merger rates depend only on $f_{\textrm{PBH}}$ and not on $f_{\mathrm{PBH \, binaries}}$ \cite{Aljaf:2024fru}.

Regarding the merger rate from binary-single interactions inside dark matter halos,  Ref.~\cite{Aljaf:2024fru}, studied the interactions of hard PBH binaries with PBHs in halos. Through such interactions, hard binaries can get harder and their merger times accelerate. However, the frequency of this process depends on the density and the dispersion velocity of halos, which in turn depends on location within each halo and the halo's evolutionary stage. As shown in~\cite{Aljaf:2024fru}, the resulting merger rates can be dominant over the direct capture rates. 

In this work, we compute the merger rate from binary-single interactions by extending the analysis of Ref.~\cite{Aljaf:2024fru}. We simulate a given halo by partitioning it in up to ten mass shells and evolve them in time starting from redshift of $z=12$ to today. This allows us to properly account for the position and time dependence of the dispersion velocity $v_{\textrm{disp}}$ of the PBHs and their density that follows $\rho_{\textrm{NFW}}$.  The evolution of orbital properties of hard PBH binaries inside the halos follows,
\begin{equation}
\label{eq:evol_a}
\frac{d a}{d t}=-\frac{G \cdot H \cdot \rho_{\mathrm{env}}(r, t)}{v_{\mathrm{disp}}^{\textrm{env}}(r, t)} \, a^2-\frac{64}{5} \frac{G^3}{c^5 \, a^3} 
\cdot f(m) \cdot F(e)
\end{equation}
\begin{eqnarray}
\label{eq:evol_e}
\frac{d e}{d t}&=&\frac{G\cdot H(r, t) \, K(r, t) \cdot \rho_{\mathrm{env}}(r, t)}{v_{\text {disp}}^{\text {env}}(r, t)} \, a \nonumber \\
&-& \frac{304}{15} \frac{G^3}{c^5 \, a^4} 
 \cdot f(m)\cdot D(e),
\end{eqnarray}
with $f(m)=(m_1+m_2)\cdot (m_1\cdot m_2)$ and $F(e)$ and $D(e)$ being functions of the eccentricity of the binary (see Appendix \ref{app:halo_properries} for their explicit form).  The $\rho_{\mathrm{env}}$ and $v_{\textrm{disp}}^{\textrm{env}}$, are to indicate that for all binaries we keep track of their environment's evolution (i.e. mass shell) from redshift of 12 to 0. 
We start our simulations with an initial sample at $z = 12$, consisting of early PBHs binaries that have survived GW mergers since their formation, and evolve them to present day $z=0$ using Eqs.~\eqref{eq:evol_a}--\eqref{eq:evol_e} for each mass shell. At each time step of the simulation, if the semi-major axis $a$ of a binary approaches zero, we consider that binary to have merged. The number of mergers at each time step is then rescaled by the number of binaries in the shell to obtain the merger rate per shell per time step. Summing over all shells gives the total merger rate per halo of a given mass at each time step (\eqref{eq:Rhalorescaled}).
To obtain the comoving PBH merger rate from binary-single interactions over the entire range of halos at a given redshift, we again use Eq.~(\ref{eq:HMF}). 

Ref.~\cite{Aljaf:2024fru} provides, for monochromatic $30 \, M_\odot$ PBHs, all details on the calculation of the total merger rate due to direct captures and binary-single interactions, by evolving only hard binaries. In this work, we extended the analysis to include both hard and soft PBH binaries and performed simulations for 40 halos with their present day masses ranging $10^4 - 10^{15} \, M_\odot$, assuming PBHs to be monochromatic with mass $m_{\rm PBH} = 30 \, M_{\odot}$ and $f_{\textrm{PBH\,binaries}}=0.5$. We also run simulations for other PBH masses with $5-80 \, M_{\odot}$ found the following rescaling formula is valid for the comoving merger rate in terms of $m_{\textrm{PBH}}$:
\begin{eqnarray}
\label{eq:R_rescaled}
R_{\textrm{binary-single}}(z, m_{\textrm{PBH}}) &=& \int_{M_{\textrm{min}}}^{M_{\textrm{max}}} R_{\textrm{binary-single}, 30}^{\text{halo}}(z, M)  \nonumber \\
&\times& s(m_{\textrm{PBH}}) \, \frac{d n}{d M} \, dM,
\end{eqnarray}
where $R_{\textrm{binary-single}}(z, m_{\textrm{PBH}})$\footnote{To test the impact of  halo mass function, we calculated this rate using the Sheth-Tormen mass function\cite{Sheth:1999su}. We found a difference within $10\%$ compared to the Press-Schechter.}
is the binary-single merger rate across halos for monochromatic PBHs of mass $m_{\textrm{PBH}}$, at redshift $z$. $R_{\textrm{binary-single}, \, 30}^{\textrm{halo}}(z, M)$ denotes the  binary-single merger rate per halo, at redshift $z$ for PBHs with a mass of $30 \, M_\odot$ in a halo of mass $M$. The rescaling function $s(m_{\textrm{PBH}})$ is,
\begin{equation}
    s(m_{\textrm{PBH}})=\delta\cdot\left(\frac{30 M_\odot }{m_{\textrm{PBH}}}\right)^{\gamma}
\label{eq:rescaling_function}    
\end{equation}
with $\gamma=0.8$ and $\delta=1$ \footnote{For PBHs with a log-normal mass distribution, we follow the same prescription of Eqs~(\ref{eq:R_rescaled}),(\ref{eq:rescaling_function})  with $\gamma=0.8$ and  $\delta = 1.25$ instead.}. This formula includes the merger rates of PBHs that would merge even in  isolation and their time to merger gets reduced due to binary single interactions.

\begin{figure}[ht!]
    \centering
    \includegraphics[width=1\linewidth]{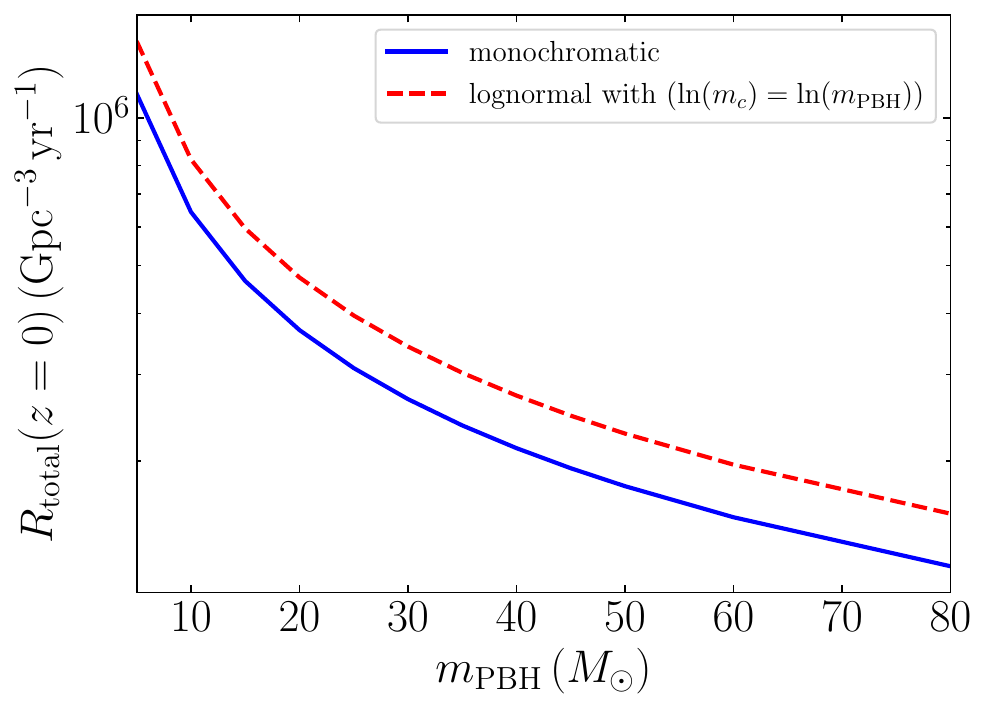}

    \caption{The total merger rate of PBH binaries at $z = 0$, due to t body captures and binary-single interactions and unperturbed binaries as a function of $m_{\textrm{PBH}}$. The blue solid line gives the rate for a monochromatic distribution. The red dashed line gives instead the rate for a log-normal mass distribution with $\mu \equiv \textrm{ln}(m_{c}) =\textrm{ln}(m_{\textrm{PBH}})$ and $\sigma = 0.6$. In all cases we assumed $f_{\textrm{PBH\, binaries}}=0.5$}
    \label{fig:Total_rate_vs_mPBH}
\end{figure}

Fig.~\ref{fig:Total_rate_vs_mPBH} depicts the total merger rate of PBH binaries from  captures, binary single interactions and unperturbed binaries at $z = 0$, as a function $m_{\textrm{PBH}}$ for the monochromatic distribution (blue solid line), i.e,

\begin{equation}
R_{\rm total} = R_{\rm captures} + R_{\rm binary-single} + R_{\rm unperturbed}
\end{equation}

Although the direct capture rate is independent of $m_{\textrm{PBH}}$,  the dependence of the total rate on $m_{\textrm{PBH}}$ originates from binary-single interactions channel unperturbed binaries channel. Their rate depends on $m_{\rm PBH}$ because both formation channels scale as $m_{\rm PBH}^{-0.8}$, leading to a higher merger rates for lower PBH masses.

The rate of PBH binary-single interactions per volume inside a halo is given by,
\begin{eqnarray}
\frac{dR_{\textrm{binary-single}}^{\text{halo}}}{dV} &=& n_{\textrm{hard-binaries}} \cdot n_{\textrm{single PBH}} \nonumber \\
&\times& \sigma_{\textrm{binary-single}} \cdot v_{\textrm{disp}}.
\label{eq:R_binary_single}
\end{eqnarray}
$n_{\textrm{single PBH}}$, is the number density of single PBHs (including PBHs in wide binaries) and $\sigma_{\textrm{binary-single}}$ is the interaction cross-section of PBH binaries with other PBHs.
The number density of hard binaries is $n_{\textrm{hard-binaries}}=f_\textrm{hard}\cdot n_{\textrm{PBH binaries}}$, where $n_{\textrm{PBH binaries}}$ is the density of PBH binaries and $f_\textrm{hard}$ the fraction of them that are hard, i.e. the PBH binaries with a semi-major $a$ below the critical value of,
\begin{equation}
a_{h}(r, t) = \frac{G \, m_\textrm{PBH}}{4v_\textrm{disp}(r, t)^2}.  
\label{eq:a_h}
\end{equation}
Only (hard) binaries with $a<a_{h}$, will merge within the Hubble time. This criterion is environment-dependent due to $v_\textrm{disp}(r, t)$, the dispersion velocity of the dark matter halo that the PBHs are in.
Eq.~(\ref{eq:a_h}) for $a_h(r, t)$ shows that $a_h$ increases linearly with the mass of $m_{\textrm{PBH}}$. 
In massive halos, the dispersion velocity $v_{\textrm{disp}}(r, t)$ is high and only a small fraction of the PBH binaries are hard. 
In fact, in the very massive halos we find that approximately, $f_{\textrm{hard}}\propto m_{\textrm{PBH}}$. However, in the smallest dark matter halos $M \sim 10^{3} \, M_{\odot}$, almost all PBH binaries are hard, i.e. $f_{\textrm{hard}}\simeq 1$.  
In Appendix~\ref{app:Hard_Binaries_fractions}, we provide further details on how the environments the PBH binaries are in, affect the ratio $f_{\textrm{hard}}$.

For a given value of semi-major axis, the cross-section for PBH binary-single interaction scales with $m_{\textrm{PBH}}$,
\begin{equation}
\sigma_{\textrm{binary-single}}= \frac{2 \pi G \cdot (3\,m_{\textrm{PBH}}) \cdot a}{v_{\textrm{disp}}^2}\propto m_{\textrm{PBH}}.
\label{eq:sigma_binary_single}
\end{equation}

Combining all those dependencies on $m_{\textrm{PBH}}$, after integrating over the volume of the halo, we find from Eq.~(\ref{eq:R_binary_single}), that in large halos, the PBH binary-single interaction rate per halo is,
\begin{eqnarray}
        R_{\textrm{binary-single}}^{\textrm{halo}} &\propto& \left( \frac{\rho_{\textrm{NFW}}}{2 m_{\textrm{PBH}}} \right) \cdot m_{\textrm{PBH}} \cdot \left( \frac{\rho_{\textrm{NFW}}}{m_{\textrm{PBH}}} \right) \cdot m_{\textrm{PBH}} \nonumber \\ 
        &\propto& \textrm{independent to } m_{\textrm{PBH}}.
\end{eqnarray}

However, in the low mass halo limit, the same PBH binary-single interaction rate per halo is approximately, 
\begin{eqnarray}
        R_{\textrm{binary-single}}^{\textrm{halo}} &\propto& \left( \frac{\rho_{\textrm{NFW}}}{2 m_{\textrm{PBH}}} \right) \cdot 
        \left( \frac{\rho_{\textrm{NFW}}}{m_{\textrm{PBH}}} \right) \cdot m_{\textrm{PBH}} \nonumber \\
        &\propto& \frac{1}{m_{\textrm{PBH}}}. 
\end{eqnarray}
This explains the total merger rate having some dependence on $m_{\textrm{PBH}}$. 

The difference between the detailed simulations and the simple analytical arguments given above, is that in Eq.~(\ref{eq:sigma_binary_single}), the PBH binary-single interaction cross-section also depends on the semi-major axis $a$, whose critical (upper) value of what defines a hard binary also depends on $m_{\textrm{PBH}}$ (see Eq.~(\ref{eq:a_h})). 

We find that the full numerical treatment to account for the halo effects on the merger rate gives an increased rate by a factor of $\simeq 40 \%$ compared to the evaluation from earlier analyses as Ref.~\cite{Ali-Haimoud:2017rtz, Raidal:2017mfl} (see Ref.~\cite{Aljaf:2025dta} for more details).

\subsection{Primordial black holes following a log-normal mass distribution}
\label{subsec:rates_log-normal}

To simulate the merger rates for the halos made of PBHs with a mass distribution function, we take the log-normal mass distribution, defined by the probability density function,
\begin{equation}
\psi(m)=\frac{1}{\sqrt{2 \pi} \, \sigma \, m} \exp \left(-\frac{\left(\ln(m) -\mu\right)^2}{2 \sigma^2}\right), 
\label{eq:MassPDF_lognormal}
\end{equation}
with $\mu =ln(m_c)$, where $m_c$ represents the median mass (in units of $M_{\odot}$) and $\sigma$ characterizes the width of the mass distribution. 

In our halo simulations for binary-single channel, when we have terms that include the mass of PBH binaries or of single PBHs as in $f(m)$, $q(m)$ and $a_h(m)$, we derive averaged values for these quantities by using the PBH mass distribution function. 

The mass term $f(m)$ in the evolution equations Eqs.~(\ref{eq:evol_a}) and (\ref{eq:evol_e}), is as follows,
\begin{eqnarray}
     \left\langle f(m)\right\rangle &=& \int_{m_{2,min}}^{m_{2,max}} \int_{m_{1,min}}^{m_{1,max}} (m_1+m_2)(m_1 \cdot m_2)  \nonumber \\ 
     &\times& \psi(m_1) \, \psi(m_2)\,dm_1  \,dm_2.
\end{eqnarray}

The black hole mass ratio $q =m_2/m_1$ of binaries, for the monochromatic case is $q= 1$. 
For the log-normal distribution, as $m_{2} < m_{1}$ we have, 
\begin{equation}
\left\langle q(m)\right\rangle = \int_{m_{1,\textrm{min}}}^{m_{1,\textrm{max}}} \frac{1}{m_1} \psi(m_1) \, dm_1 
\int_{m_{2,\textrm{min}}}^{m_{1}} m_2 \, \psi(m_2) \, dm_2.
\end{equation}
This term affects the eccentricity growth rate $K(r,t,m)$ in  Eqs.~(\ref{eq:evol_a}) and (\ref{eq:evol_e}).

The fraction of hard and soft binaries for a given halo mass and location within it (the spherical shell of the halo), depends on the PBH mass distribution, through its impact on the critical value $a_{h}$,
\begin{equation}
\label{eq:a_h_lognormal}
a_{h}(r, t) = \frac{G \, m_1\cdot m_2}{4v_\textrm{disp}(r, t)^2 \, m_3}.  
\end{equation}
$m_1$ and $m_2$ are the masses of the PBH binary and $m_3$ is the mass of the interacting single PBH.
Thus, for a log-normal PBH mass distribution,
\begin{eqnarray}
     \left\langle a_{h}(m) \right\rangle&=&\frac{G}{4 v_{\textrm{disp}}^2} \int_{m_{3,min}}^{m_{3,max}} \int_{m_{2,min}}^{m_{2,max}} \int_{m_{1,min}}^{m_{1,max}} \frac{m_1\cdot m_2}{m_3}  \nonumber \\
     &\times& \psi(m_1) \, \psi(m_2) \, \psi(m_3)\,dm_1  \,dm_2 \,dm_3.
\end{eqnarray}

In our simulations, we sample and evolve a large enough number of binaries to properly study the rare stochastic processes of direct captures and binary-single interactions and unperturbed binaries. Thus, we need to properly rescale the number of studied binaries to the actual number of binaries existing in each halo mass shell $i$, at any given time $t$ in the evolution of each simulated halo. For a PBH mass-distribution the actual number of PBH binaries $N_{\textrm{PBH binaries},\, i, \, t}$ in a shell indexed by $i$ at time $t$ is approximated as, 
\begin{eqnarray}
    N_{\textrm{PBH binaries}, \, i, \, t}=\frac{M_{i, \, t}}{4\left\langle m \right\rangle}.
\label{eq:NPBHbinaries}    
\end{eqnarray}
$M_{i, \, t}$, is the dark matter mass in PBHs in the shell $i$ at time $t$.
The mean PBH mass $\left\langle m \right\rangle$ is,
\begin{equation}
     \left\langle m \right\rangle=\int_{m_{min}}^{m_{max}}  m  \, \psi(m) \, dm.
\end{equation}
The factor of 1/4 in Eq.~(\ref{eq:NPBHbinaries}), is due to the fact that we assume $f_{\textrm{PBH binaries}}=0.5$, and each binary is made up of two PBHs.
The properly rescaled merger rates per halo is,
\begin{eqnarray}
    R_{\textrm{binary-single}}^{\textrm{halo}} &=& \sum_{i=1}^{N_{\textrm{shells}}} \sum_{t=0}^{t_{\textrm{look}}} \frac{N_{\textrm{PBH binaries}, \, i, \, t}}{N_{\textrm{sample},\, i,\, t}} \cdot \frac{N_{\textrm{merger},\, i,\, t}}{\Delta t}. \nonumber \\
    &&
\label{eq:Rhalorescaled}    
\end{eqnarray}
In Eq.~(\ref{eq:Rhalorescaled}), $t_{\textrm{look}}$ is the look back time, for $N_{\textrm{shells}}$ number of spherical shells that we divide the dark matter halo.
Also, $N_{\textrm{sample},i,t}$ is the number of PBH binaries that are sampled at the shell $i$ at a given timestep associated with time $t$. 
$N_{\textrm{PBH binaries}, \, i, \, t}$, is the true number of PBH binaries in the shell $i$ at time $t$. 
$N_{\textrm{merger},i,t}$ is the number of the binaries that merged in our simulation within the mass shell $i$, at the timestep associated with time $t$. 
Our timesteps have a duration of $\Delta t$. 
For more details on the simulations see Refs.~\cite{Aljaf:2024fru, Aljaf:2025dta}. 

Additionally, to the above mentioned quantities, the initial distribution functions of the PBH binaries' orbital parameters $a_{0}$ (for the semi-major axis) and $e_{0}$ (for the eccentricity), depend on the PBH mass distribution. 
These initial distributions of the PBH binaries' properties, 
originate from the initial early universe conditions at the creation of PBHs. In our simulations, for each of our binaries, we randomly pick its initial values of semi-major axis and eccentricity from these initial distribution functions. We then evolve $a$ and $e$, based on Eqs.~(\ref{eq:evol_a}) and~(\ref{eq:evol_e}).
In Appendix~\ref{app:BBHs_Initial_Semimajor_Distribution}, we provide further details on how we calculate the initial distribution functions of the PBH binaries' orbital properties for the continuous log-normal PBH mass distribution.

In Fig.~\ref{fig:Total_rate_vs_mPBH}, for log-normal distributions with $\sigma=0.6$, we show the total merger rate as a function of $m_c$ (red dashed line). A choice of $\sigma \ll 0.6$, would give a mass distribution effectively identical to a monochromatic one, given the current mass resolution capacity of LVK. Instead, values of $\sigma \gg 0.6$, will result in very wide mass distributions for the PBH binaries. Given the limited number of observed binaries by LVK, such a PBH component to the data is very difficult to separate from a more conventional power-law distribution in the masses of the black holes. 
The log-normal mass-distribution provides a relatively higher rate than the monochromatic one. 

\section{Simulating the detectability of primordial black hole binaries and setting limits}
\label{sec:BBHs_limits}

In order to set limits on a PBH component to the LVK O1-O3 observations, we  
implement the methodology of Ref.~\cite{Bouhaddouti:2024ena}. 
To simulate the LVK observations, we generate merging PBH binaries with masses $m_{1}$ and $m_{2}$, following the relevant PBH mass-distribution (monochromatic or log-normal). 
These merging PBH binaries, have the redshift distribution $R_{\textrm{total}}(z)$ described earlier. For all these merging binaries we evaluate their signal-to-noise ratio (SNR) and retain the binaries with a SNR$>8$. 
We also generate a population of ``conventional'' black hole binaries with $m_{1}$ following a power-law distribution: $dN/dm_{1} \propto m_1^{-\alpha}$ and the ratio of the secondary mass over the primary mass $q=m_2/m_1$ following a power-law  distribution as well, $dN/dq \propto q^\beta$. These binaries have a redshift distribution that resembles that of star formation \cite{Madau:2014bja}. Their comoving rate density is 
$dR/d(z+1) \propto (1+z)^\kappa$, with $\kappa = 2.9$.
Our assumptions on this conventional population of black hole binaries follow the same prescription of studying the LVK observations as Ref.~\cite{PhysRevX.13.011048} (see also~\cite{Kovetz:2016kpi}). 
We refer to this population of black holes as a power-law population or just ``PL''. 
We retain the conventional PL population of black hole binaries with SNR$>8$. 
We then compare the combined population of merging PBH binaries and black hole binaries from a PL population to the $m_{1}$- and $q$-histograms of the LVK O1-O3 observations, presented in Ref.~\cite{Bouhaddouti:2024ena}.
For a given choice of PBH mass distribution, we marginalize over the values of $\alpha$, $\beta$ and the normalizations of the PBH and PL components. 

We consider 15 bins of redshift, with values  from z=0 to z=0.75 in increments of 0.05. Above a redshift of 0.75 binaries are not detectable using the O3 noise curve provided by the LVK collaboration \cite{LIGOnoise}. 
The number of mergers in a bin with edges $z_1$ and $z_2$ is,
\begin{equation}
    N_{\textrm{merger}}(z_1,z_2)= 4\pi \int_{z_1}^{z_2} \frac{c \cdot \chi(z)^2 \cdot R_{\textrm{total}}(z)}{(1+z)H(z)} \, dz,
\end{equation}
with $\chi(z)$ the comoving distance, $R_{\textrm{total}}(z)$ the PBH comoving merger rate and $H(z)$ the Hubble expansion parameter.

In the LVK observations, there is a generic uncertainty on the derived masses of the detected black hole binaries.
For a monochromatic PBH component, we approximate that error by convolving with a Gaussian that has a mass-dependent width (see table \RNum{4} in \cite{Bouhaddouti:2024ena} for further details). 
Thus, even for an underlying monochromatic PBH population, the generated PBH binaries have masses $m_{1}$ and $m_{2}$ with $m_{2} \leq m_{1}$. For a given underlying $m_{\textrm{PBH}}$ we draw the $m_{1}$ and $m_{2}$ masses from the relevant Gaussian density function, with a width dependent on $m_{\textrm{PBH}}$. For instance for $m_{\textrm{PBH}} = 10 \, M_{\odot}$, the Gaussian has a $\pm 1 \sigma$-width of $\pm 2.1 \, M_{\odot}$, while for $m_{\textrm{PBH}} = 30 \, M_{\odot}$, the $\pm 1 \sigma = \pm 5.1 M_{\odot}$ (see Ref.~\cite{Bouhaddouti:2024ena} for more details). 
For the case of the log-normal PBH mass distribution, we first convolve the relevant underlying distribution with a Gaussian whose width depends on the value of the underlying PBH mass.
A log-normal distribution predicts PBHs with a wide range of masses. A $10\, M_{\odot}$ PBH will have its mass measured by LVK with a different uncertainty than a $60\, M_{\odot}$ PBH. We account for that mass-dependency in our simulated events. 
We subsequently draw the $m_{1}$ and $m_{2}$ values from that final distribution.

In deriving the best-fit and the 95$\%$ upper limit normalization of the PBH component, for a given choice of PBH mass, we marginalize over the values of $\alpha$, $\beta$ describing the PL black hole binaries population. 
We perform a $\chi^2$ fit of the combined PBH and PL populations to the  
the LVK $m_{1}$- and $q$-histograms of Ref.~\cite{Bouhaddouti:2024ena}.

The simulated $R_{\textrm{total}}(z)$ for the PBH population assumes $f_{\textrm{PBH}}=1$ and $f_{\mathrm{PBH \, binaries}} = 0.5$, that once fitting to the data gets significantly suppressed.
The normalization factor of the PBH population is proportional to $f_{\textrm{PBH}}^{2}$, but scales in a less trivial manner with $f_{\mathrm{PBH \, binaries}}$, relevant for the PBH merges induced by binary-single interactions.

It is important to note that due to the small LVK detected merger events, 
our current statistical errors are quite large. 
In addition, LIGO's O3 noise curves are time dependent, a piece of information not directly accessible to us. We used the noise curves of Ref.~\cite{LIGOnoise}, for the entire observing run time. 

The total merger rate's $R_{\textrm{total}}$ relation to $f_{\textrm{PBH}}$ and $f_{\mathrm{PBH \, binaries}}$ is,
\begin{equation}
R_{\rm total} = f_{\rm PBH}^{2} \cdot \Big(
      R_{\rm captures} 
      + g\cdot \big( R_{\rm binary-single} + R_{\rm unpert} \big)
\Big)
\end{equation}
with $g=2 \cdot f_{\textrm{PBH \, binaries}}$. Note that the factor of 2 is because the $R_{\textrm{binary-single}}$ and $R_{\rm unpert}$ components is evaluated for $f_{\mathrm{PBH \, binaries}} = 0.5$.

Fitting the amplitude of the PBH component (i.e.  its total comoving rate) to the LVK observations we directly probe $f_{\textrm{PBH}}^{2}$, which we refer to as $f_{\textrm{PBH, LVK}}^{2}$. 
For any value of $f_{\mathrm{PBH \, binaries}}$, we get for $f_{\textrm{PBH}}$,
\begin{equation}\label{f_PBH}
f_{\textrm{PBH}} = \left( {\frac{f_{\textrm{PBH, LVK}}^2 \cdot R_{\textrm{total}}}{R_{\textrm{captures}} +g\cdot (R_{\textrm{binary-single}}+R_{\rm unpert})}}\right)^{1/2}.
\end{equation}

\begin{figure}[ht!]
    \includegraphics[width=1.018\linewidth]{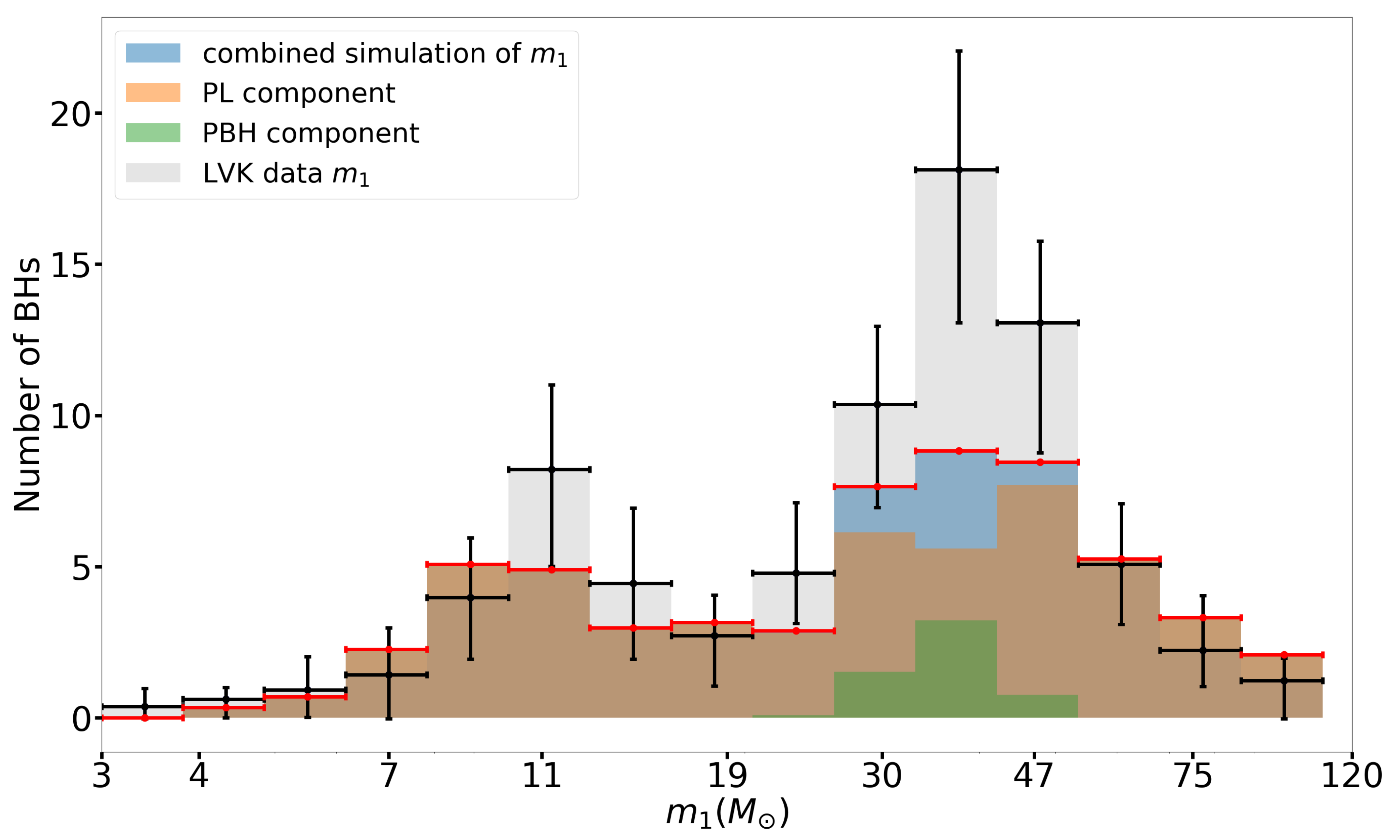}\\
    \vspace{0.1in}
    \includegraphics[width=1\linewidth]{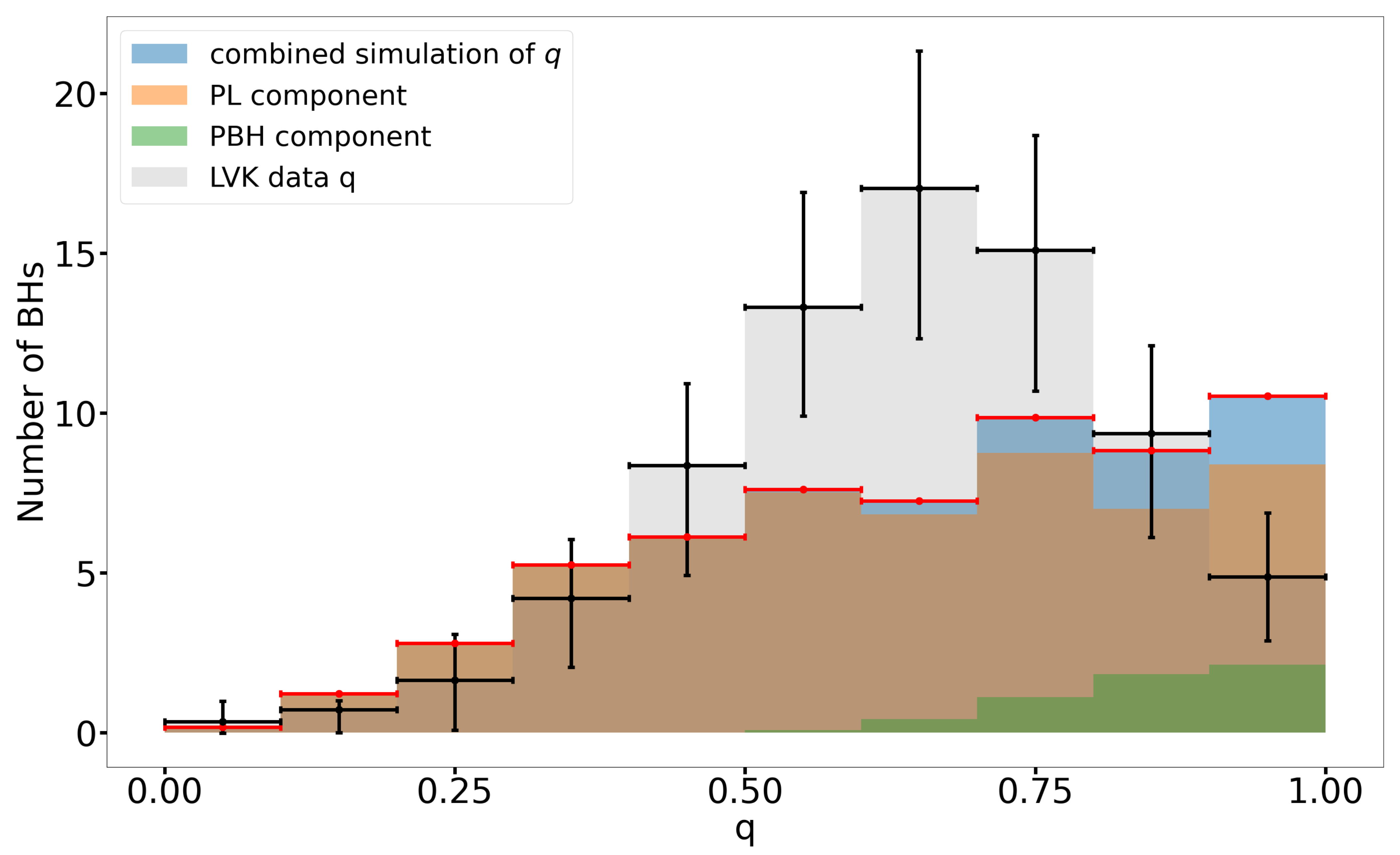}
    \caption{Normalized $m_{1}$-histogram (top) and $q$-histogram (bottom) of the simulated (in blue) and detected by LVK (in gray) black hole binaries with a SNR> 8. The simulated binaries are a combination of a dominant PL population (in brown) and a population of PBH binaries (in green). See text for further details.}
    \label{fig:m1_and_q_histrohram_35Msun_monochromatic}
\end{figure}

In Fig.~\ref{fig:m1_and_q_histrohram_35Msun_monochromatic}, we give the $m_{1}$-histogram (top panel) and the $q$-histogram (bottom panel) of simulated black hole binaries detectable with a SNR $>8$ at O3 run sensitivity. 
The underlying  simulation includes the PL populations with $\alpha = 3.44$ and $\beta = -1$. 
The PBH component is for $m_{\textrm{PBH}} = 35 \, M_{\odot}$, with $f_{\textrm{PBH}} = 1.1 \times 10^{-3}$. Superposed to that, is the $m_{1}$-histogram of the LVK observed black hole binaries with Poisson error-bars (see Ref.~\cite{Bouhaddouti:2024ena}). 
The normalizations of the two populations and the values of $\alpha$ and $\beta$ are chosen in order for their sum to provide a best fit to the \textit{combined} $m_{1}$- and $q$-histograms. 
Since a prominent PBH population, would result in many binaries with $q\simeq 1$, which is not supported by the LVK observations the normalization of the PBH component is smaller than what would have been derived using only the $m_{1}$-histogram. 
We provide examples of $m_{1}$- and $q$-histograms for the monochromatic PBH population for increased values of $f_{\textrm{PBH}}$, along with an example of best fit results for a PBH population following a log-normal distribution in Appendix~\ref{app:m1_and_q_histograms}.

The 95\% upper limits are obtained by first fitting the LVK histograms with a PL population of black hole binaries, without a PBH component. That gives a best fit $\chi^{2}$ of 33.46, for $\alpha = 3.44$ and $\beta = -1$. Then to get an upper limit on the PBH population we find the maximum normalization to the PBH component for which we get a $\chi^{2} = 33.44 +2.71 = 36.15$.

\begin{figure*}[ht!]
    \centering
\includegraphics[width=0.9\linewidth]{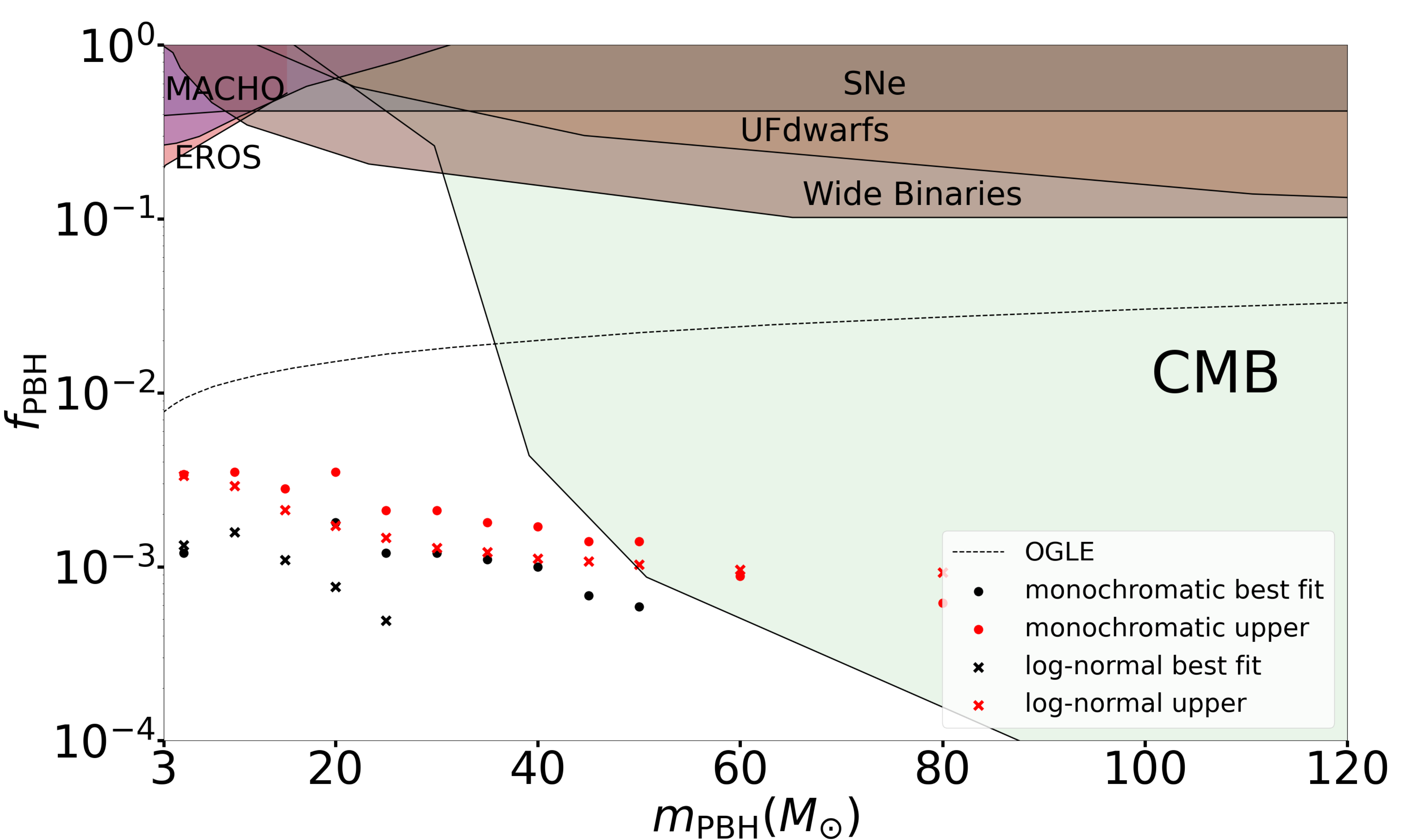}
    \caption{Our results on $f_{\textrm{PBH}}$ as a function PBH mass, assuming $f_{\mathrm{PBH \, binaries}}$=0.5. Dots are for a  monochromatic distribution, while ``x'' symbols for a log-normal distribution where $m_{\textrm{PBH}} = m_{c} = e^{\mu}$ of Eq.~(\ref{eq:MassPDF_lognormal}). Blue symbols represent best fit values for $f_{\textrm{PBH}}$, i.e. there is statistical preference for a PBH contribution to the LVK observations. Red symbols are for the $95\%$ upper limits on $f_{\textrm{PBH}}$. We also provide for comparison upper limits from other probes on PBHs. See text for more details.}
    \label{fig:PBH_lims}
\end{figure*}

We obtain two sets of values for $f_{\textrm{PBH}}$ and $f_{\mathrm{PBH \, binaries}}$. One set is for the best fit $\chi^2$ PBH normalization, and the other is the $95\%$ upper limit on $f_{\textrm{PBH}}$. In Table~\ref{table:chi_sqr_PP_results}, we give these these values of $f_{\textrm{PBH}}$, fixing $f_{\mathrm{PBH \, binaries}} = 0.5$. In that table, our second and third columns give our results for the monochromatic case, while the fourth and fifth the log-normal distribution.

\begin{table}[th!]
\centering
\begin{tabular}{|c | c | c | c | c | } 
 \hline
 $m_{\textrm{PBH}}$ & $f_{\textrm{PBH, mon.}}^{\textrm{best-fit}}$  & $f_{\textrm{PBH, mon.}}^{\textrm{95\% upper}}$ & $f_{\textrm{PBH, log-n.}}^{\textrm{best-fit}}$  & $f_{\textrm{PBH, log-n.}}^{\textrm{95\% upper}}$\\ [0.5ex] 
 ($M_\odot$) & ($\times 10^{-3}$) & ($\times 10^{-3}$) & ($\times 10^{-3}$) & ($\times 10^{-3}$)\\ [0.5ex] 
 \hline
 $5$ & $1.2$ & $3.4$ & 1.3 & 3.3 \\ 
 \hline
 $10$ & $0$ & $3.5$ & 1.6 & 2.9 \\ 
 \hline
 $15$ & $0$ & $2.8$ & 1.1 & 2.1 \\ 
 \hline
 $20$ & $1.8$ & $3.5$ & 0.76 & 1.7\\ 
 \hline
  $25$ & $1.2$ & $2.1$ & 0.49 & 1.5\\ 
 \hline
  $30$ & $1.2$ & $2.1$ & 0 & 1.3\\ 
 \hline
  $35$ & $1.1$ & $1.8$ & 0 &1.2\\ 
 \hline
  $40$ & $1.0$ & $1.7$ & 0 &1.1\\ 
 \hline
  $45$ & $0.68$ & $1.4$ & 0 &1.1\\ 
 \hline
  $50$ & $0.59$ & $1.4$ & 0 &1.0\\ 
 \hline
  $60$ & $0$ & $0.89$ & 0 &0.96\\ 
 \hline
  $80$ & $0$ & $0.55$ & 0 &0.93\\ 
 \hline
\end{tabular}
\caption{Obtained values on $f_{\textrm{PBH}}$ for $f_{\mathrm{PBH \, binaries}} = 0.5$.
First column gives the mass of a monochromatic PBH population with mass $m_{\textrm{PBH}}$, or the value of $m_{c}$ for a log-normal distribution. 
Second and third columns give the $f_{\textrm{PBH}}$, best-fit ($f_{\textrm{PBH, mon.}}^{\textrm{best-fit}}$) and $95\%$ upper limit value ($f_{\textrm{PBH, mon.}}^{\textrm{95\% upper}}$) respectively, for an underlying monochromatic PBH mass-distribution. Similarly, the fourth and fifth columns give the $f_{\textrm{PBH}}$, best-fit ($f_{\textrm{PBH, log-n.}}^{\textrm{best-fit}}$) and $95\%$ upper limit ($f_{\textrm{PBH, log-n.}}^{\textrm{95\% upper}}$) respectively, for an underlying log-normal PBH mass-distribution. For the case where a PBH component provides a better fit to the PL only population of black hole binaries, we get $f_{\textrm{PBH}}^{\textrm{best-fit}} > 0$ (second or fourth column).}. 
\label{table:chi_sqr_PP_results}
\end{table}

The LVK $m_{1}$-histogram shown (in light gray) in Fig.~\ref{fig:m1_and_q_histrohram_35Msun_monochromatic} has two peaks. One approximately at $12 \, M_\odot$ and the other at $35 \, M_\odot$. We obtain best-fit values for a PBH component with $m_{\textrm{PBH}}$ around the second peak (see Table~\ref{table:chi_sqr_PP_results}, second column). However, for $m_{\textrm{PBH}} = 10 \, M_\odot$ and $15 \, M_\odot$ we get $f_{\textrm{PBH, best-fit}}=0$.
While a PBH component at $10 \, M_\odot$ or $15 \, M_\odot$ improves the $m_{1}$-histogram's fit around the first peak, it also decreases the normalization of the PL population and its contribution to the LVK data in all other bins, including the second peak around $35 \, M_\odot$. This leads to a worse fit overall. 
On the other hand, for the log-normal PBH distribution we get a preference for PBH component only for $m_{c} < 30 \, M_{\odot}$. 

In Fig.~\ref{fig:PBH_lims}, we plot our results on $f_{\textrm{PBH}}$ for a monochromatic distribution of PBHs (dots) as well as a log-normal distribution of PBHs (``x'' symbols). This figure includes both the values of $f_{\textrm{PBH}}$ associated to the best fit to the LVK data (blue dots and ``x'') and the $95\%$ upper limits on $f_{\textrm{PBH}}$ (red dots and ``x'').
In addition we show already established limits form Refs.~\cite{Zumalac_rregui_2018, Brandt_2016, Monroy_Rodr_guez_2014, Mr_z_2024, Mr_z_2024_2, Serpico_2020, Tisserand_2007, Alcock_2001,Carr_2021} (limits obtained from \cite{Green:2020jor, PBH_bounds_zenodo}. The limits from OGLE are depicted with a dashed line as they are susceptible to significant systematics as it has been recently pointed out in Ref.~\cite{Green:2025dut}.
Our limits are roughly consistent to the predictions of Ref.~\cite{Raidal:2017mfl}. We rely on the sample of O1-O3 observations and thus our limits are tighter than that of the O1-O2 sample.

\begin{figure}
    \centering
    \includegraphics[width=0.9\linewidth]{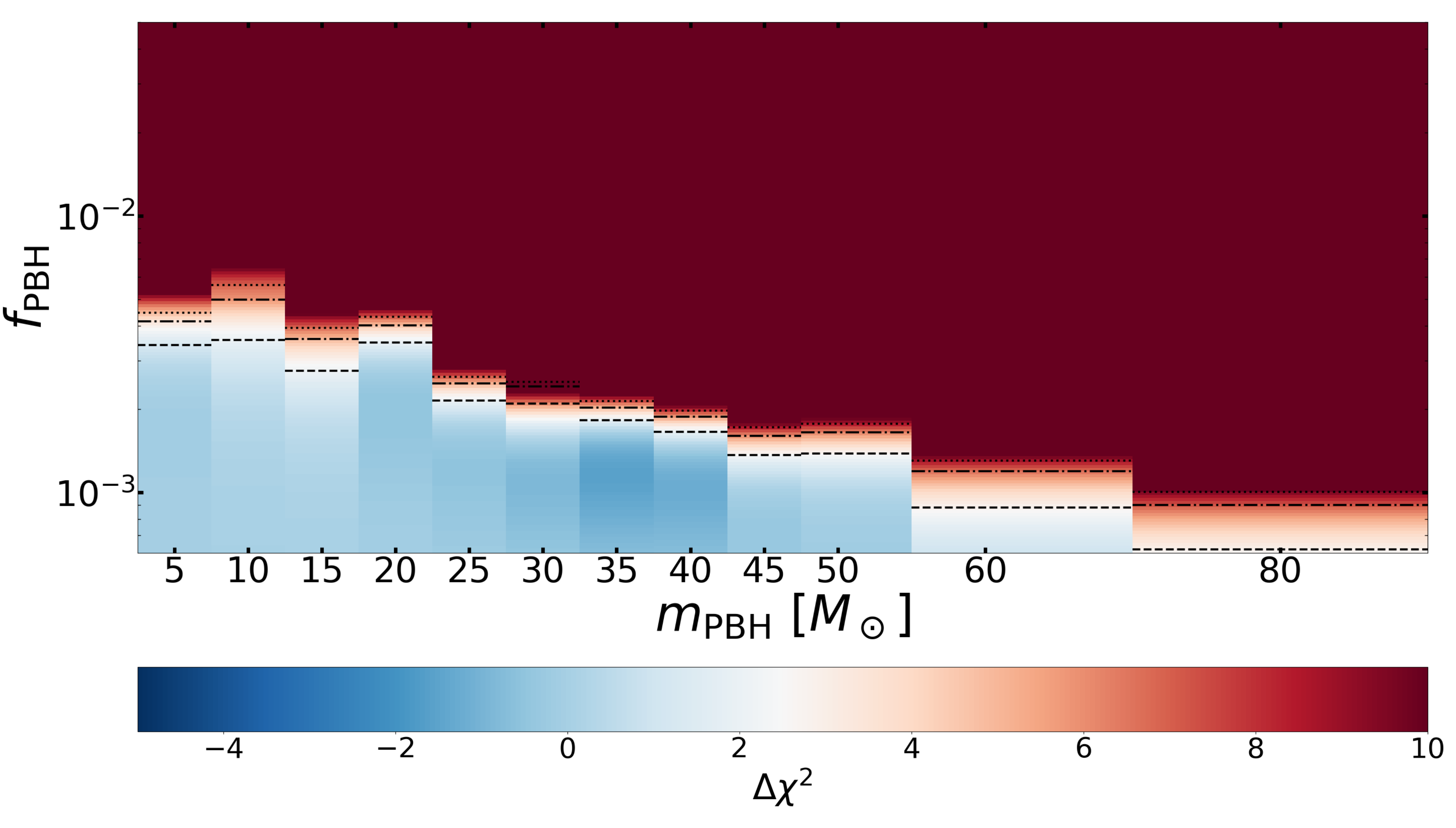}
    \includegraphics[width=0.9\linewidth]{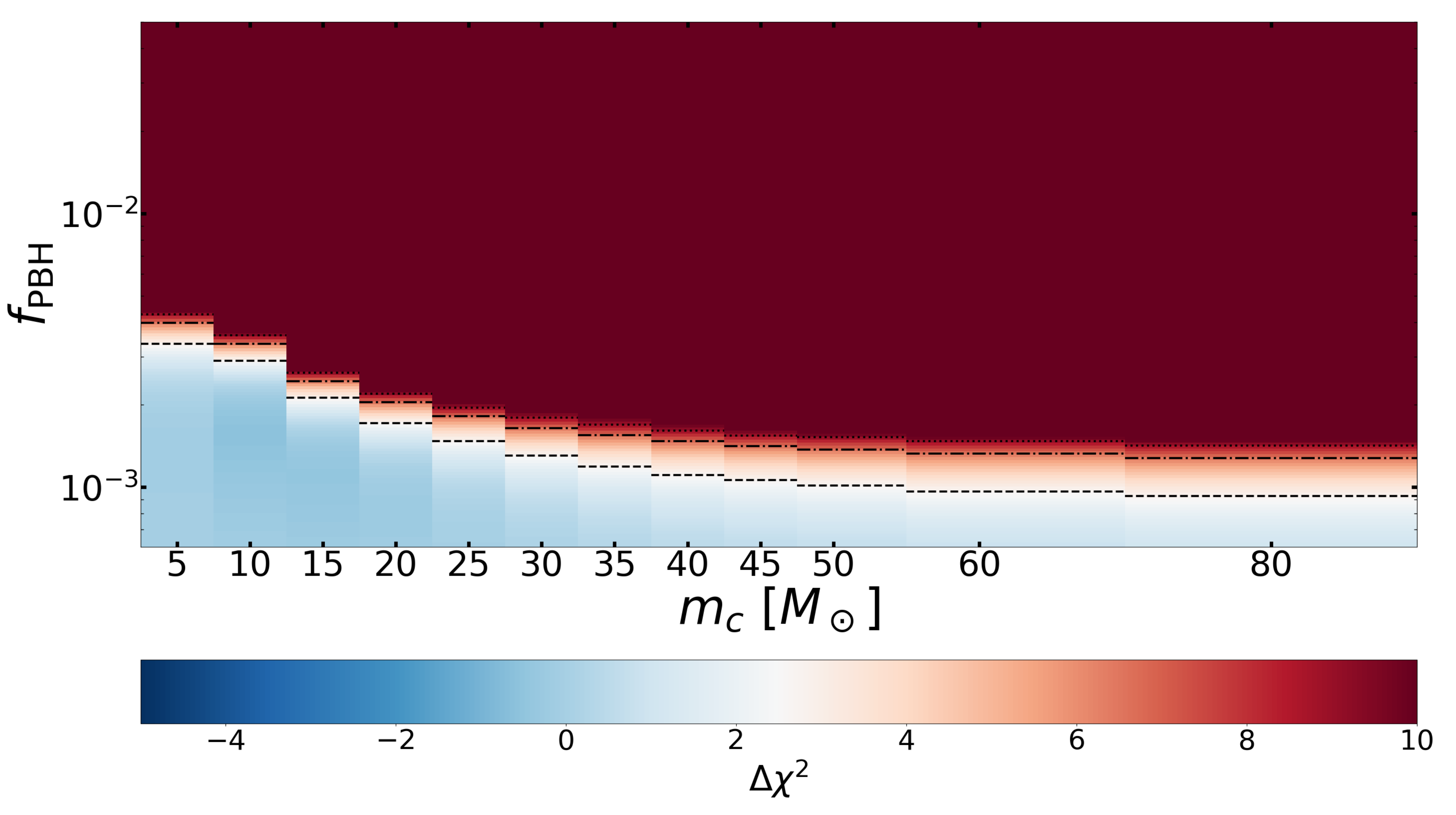}
    \caption{We present here the $\chi^2$ fit for different choices of $f_{\textrm{PBH}}$ at different masses for the monochromatic (top plot) and the log-normal (bottom plot) PBH distribution. The color gradient indicates the difference between the $\chi^2$ values using a power-law + PBH component and the best fit $\chi^2$ value of the power-law component only. The black dashed, dotted-dashed and dotted lines give the region of parameter space that is within $95\%$, $99.5\%$ and $99.87\%$ upper limits respectively. See text for further details.}
    \label{fig:chi_sqr_map}
\end{figure}

Fig.~\ref{fig:chi_sqr_map}, shows the difference in the $\chi^2$ fit values of a PL + PBH population
and a PL population of black holes only.
As explained earlier, for the PL only population we get a best fit of $\chi^2 = 33.44$. 
When we study the PL+PBH case for a given value of $f_{\textrm{PBH}}$ in each case we marginalize over the properties of the PL population (values of power-laws $\alpha$, $\beta$ and overall normalization). We also fix here $f_{\mathrm{PBH \, binaries}} = 0.5$.
We show our results for the same 12 masses of Table~\ref{table:chi_sqr_PP_results}.
White regions are for a $\Delta \chi^2=0$, i.e. $\chi^{2}= 33.44$. Blue regions are for $\Delta \chi^2 < 0$, measuring the improvement in the $\chi^{2}$-value by having a PBH population. Instead, the red regions for $\Delta \chi^2 > 0$ are used to set upper limits on the PBH population's amplitude. We assume a monochromatic PBH population for the top panel and for the bottom panel a log-normal PBH mass-distribution with $\mu = \textrm{ln}(m_{c})$ and $\sigma = 0.6$. 

For masses of $30 - 40  \, M_\odot$ there is a small improvement on the $\chi^2$ value, by $\Delta \chi^{2}\simeq 2$, compared to the best-fit PL population only case. This better fit is for $f_{\textrm{PBH}} \simeq 1-2 \times 10^{-3}$ (see Fig.~\ref{fig:chi_sqr_map}). The $95\%$, $99.5\%$ and $99.87\%$ upper limits on $f_{\textrm{PBH}}$, are also represented in Fig.~\ref{fig:chi_sqr_map}. They are associated with a $\Delta \chi^2$ of 2.71, 6.63 and 9.00 respectively. 
We have also tested a log-likelihood approach to studying the LIGO-observations represented in  Figure~\ref{fig:m1_and_q_histrohram_35Msun_monochromatic}, relying on \texttt{emcee} \cite{2013PASP..125..306F}. The results are effectively the same to those presented here. Moreover, our results regarding the power-law component are in good agreement with the results of \cite{Golomb:2023vxm} and on the PBH abundance with those of Ref.~\cite{Hall:2020daa}.  

\begin{figure}[ht!]
    \centering
    \includegraphics[width=0.99\linewidth]{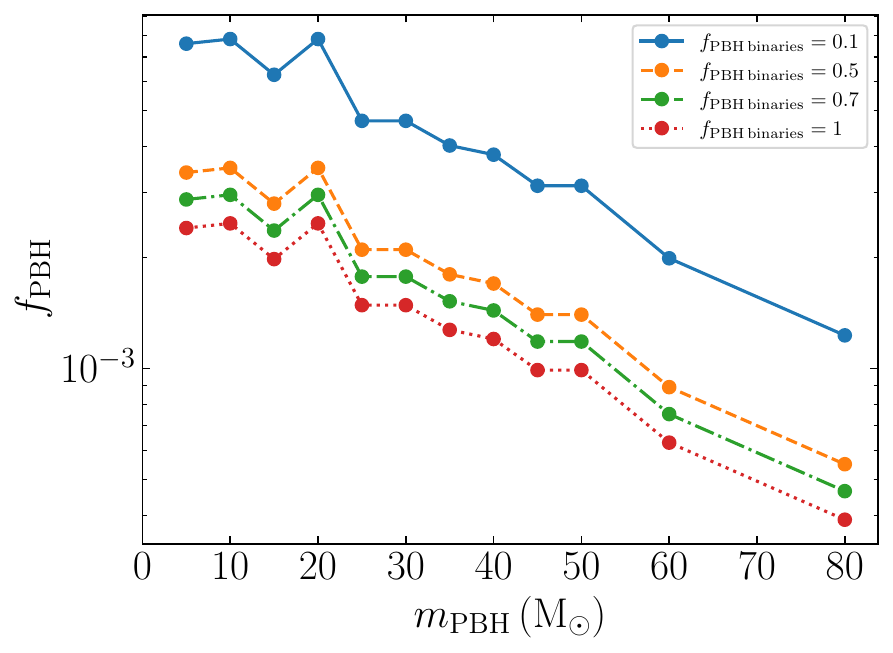}
    \includegraphics[width=0.99\linewidth]{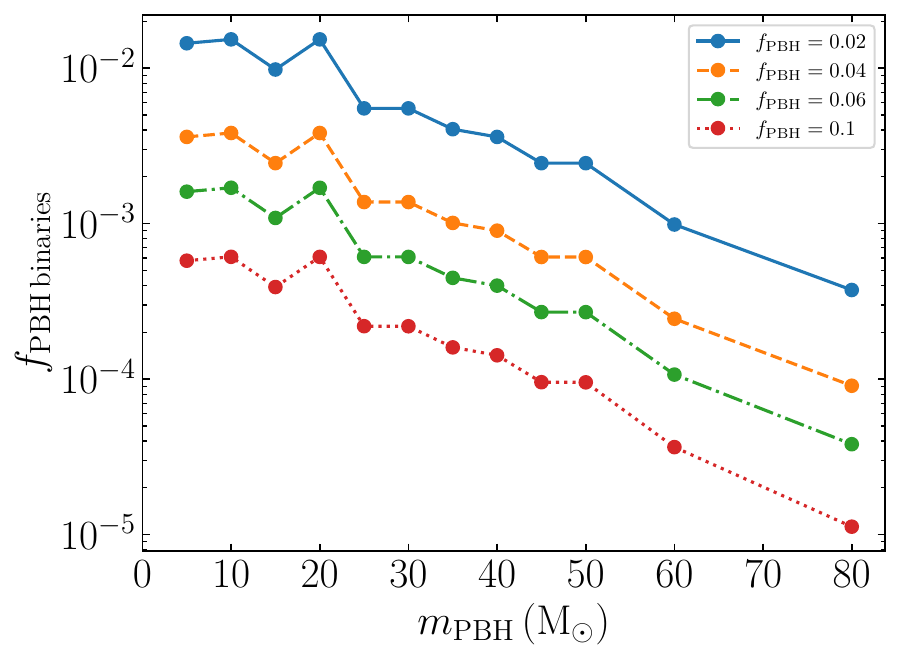}
    \caption{\textit{Top panel:} for the monochromatic PBH mass-distribution, we show how the $95\%$ upper bound on $f_{\textrm{PBH}}$ changes with $m_{\textrm{PBH}}$. We evaluate these limits by fixing $f_{\rm{PBH \, binaries}}$ to $(0.1, 0.3, 0.7, 1)$. \textit{Bottom panel:} the $95\%$ upper limits on $f_{\rm{PBH \, binaries}}$ as a function of $m_{\textrm{PBH}}$, fixing for each line $f_{\textrm{PBH}}$ to $(0.02,0.04 , 0.06,0.1)$.}.
    \label{fig:fPBH_and_fPBHbinary}
\end{figure}

In Fig.~\ref{fig:fPBH_and_fPBHbinary}, we present the $95\%$ upper limits on $f_{\rm{PBH}}$ as a function of the monochromatic PBH mass $m_{\textrm{PBH}}$, for different choices of $f_{\rm{PBH \, binaries}}$ (\textit{top} panel). As we explore a range of values for $f_{\rm{PBH \, binaries}}$ of 0.1, 0.5, 0.7 and 1 respectively, we observe that the enhancement in $f_{\rm{PBH \, binaries}}$ suppresses $f_{\textrm{PBH}}$.
This is due to the dominant contribution of PBH binary-single interactions to the total PBH merger rate.
The calculation is performed at redshift $z=0$ using Eq.(\ref{f_PBH}), with the upper limits on $f_{\rm{PBH}}$ presented in Table \ref{table:chi_sqr_PP_results} for $f_{\rm{PBH \, binaries}} = 0.5$. Moreover, at the \textit{bottom} panel of Fig.~\ref{fig:fPBH_and_fPBHbinary}, we show the $95\%$ upper limits on $f_{\rm{PBH \, binaries}}$ as a function of $m_{\textrm{PBH}}$ and how possible choices of $f_{\rm{PBH}}$ affects it. We observe a similar behavior as $f_{\rm{PBH \, binaries}}$ is being suppressed by enhancing $f_{\rm{PBH}}$. We show results for $f_{\rm{PBH}}$  of 0.02, 0.04 , 0.06 and 0.1. 

For the log-normal PBH mass-distribution, we find similar limits on $f_{\textrm{PBH}}$ to those derived for the monochromatic case. This is shown in Fig.~\ref{fig:PBH_lims}, (compare dots with ``x''s) and in the results of 
Table~\ref{table:chi_sqr_PP_results}. 
One distinction between the results for the monochromatic and the log-normal distributions, is that the mass range for which we get a (small statistical) preference for a PBH component is moved at lower masses for the log-normal assumption. 

\begin{figure}
    \centering
    \includegraphics[width=0.99\linewidth]{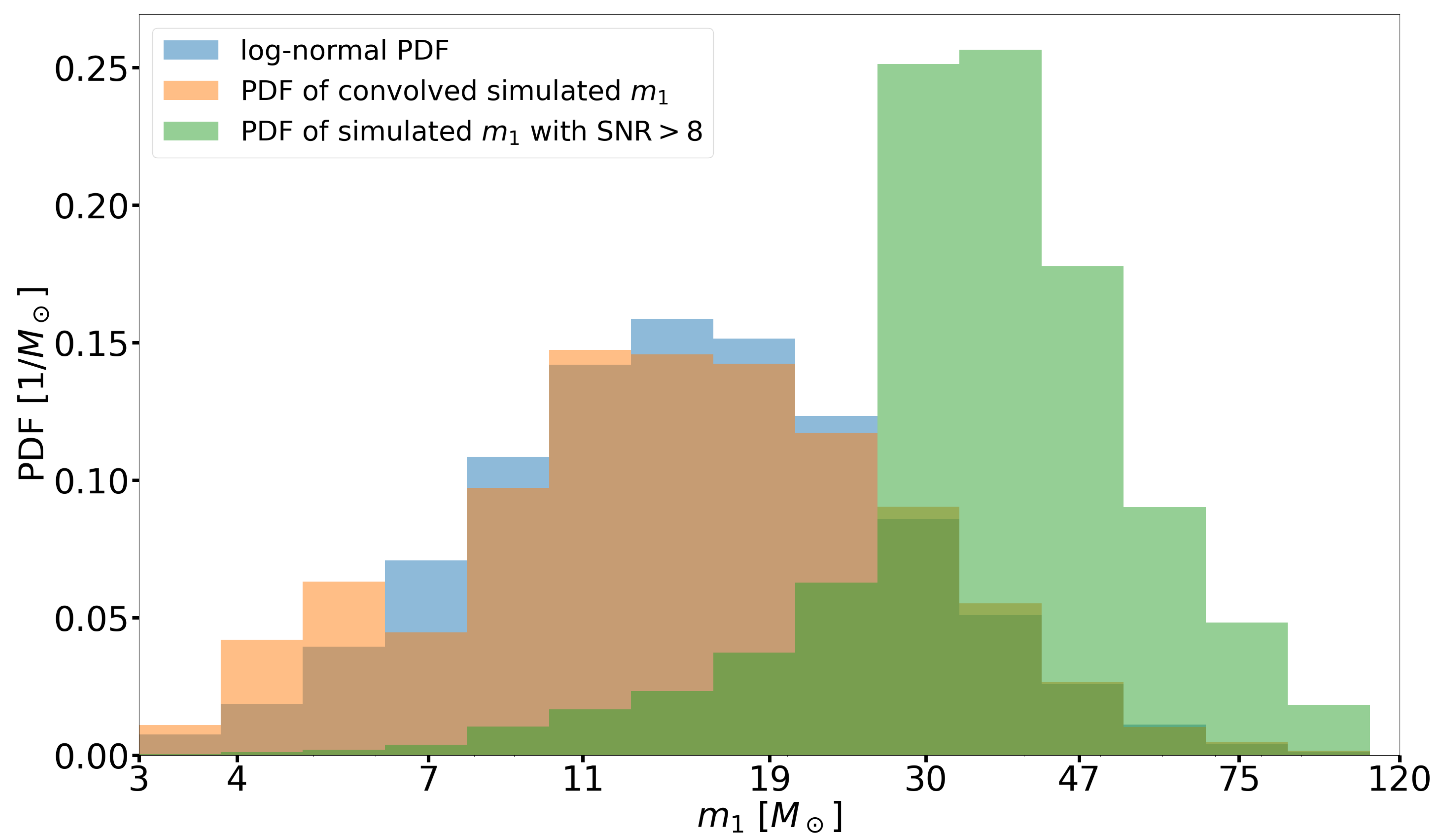}
    \caption{Histograms showing the PDFs of the mass of PBHs with a log-normal mass distribution with a peak at $15 M_\odot$ and $\sigma=0.6$ (in blue), the convolved log-normal PDF with Gaussians to take into account the uncertainty on the derived masses (in orange). In green we show the PDF of the events simulated to have a SNR$>8$, drawn from the log-normal distribution that is convolved with the Gaussians (i.e. the orange PDF).}
    \label{fig:lognorm hists}
\end{figure}

The selection of simulated events with $\textrm{SNR}>8$ changes significantly the shape of the original distribution. In Fig.~\ref{fig:lognorm hists}, we plot the normalized PBH mass distribution, i.e. probability density function (PDF). Our blue histogram is the true underlying log-normal distribution for $m_{c} = 15 \, M_{\odot}$, giving a peak of that distribution around the same value of mass. Our orange histogram give the PDF after convolving the original log-normal distribution with the Gaussian of mass-dependent width to account for the mass resolution of LVK. Finally, the green histogram 
shows the PDF on $m_{1}$ after retaining only the PBH binaries with a SNR$>8$. 
The peak of the final observable population is shifted to a value of $35 M_\odot$. 
From the $\chi^2$ map of Fig.~\ref{fig:chi_sqr_map}, derived for the log-normal mass-distribution (bottom panel), we see that such a distribution with a peak near $m_{c} = 15 \, M_\odot$ is better than a power-law fit alone. 
This is due to the effect of selecting binaries with a SNR>8, which shifts the peak closer $40 \, M_\odot$ (see figure 7), around which there is a significant peak present in the LVK $m_{1}$-histogram.

\section{Discussion and conclusions}
\label{sec:conclusions}
In this paper we have explored the impact of the recent LVK observing runs (O1-O3) on the abundance of PBHs. We present limits for monochromatic PBHs and for PBHs following a log-normal mass-distribution. Given the limited mass resolution of the LVK observations, for the log-normal distribution, we choose a width that can be distinguished form the monochromatic case, while also would not be confused to being part of a population of black hole binaries that would roughly follow a power-law distribution in their mass.

We evaluate the PBH merger rates by including all relevant binary formation and evolution channels: (i) binaries formed through direct gravitational-wave captures during close encounters inside dark-matter halos; (ii) mergers of PBH binaries whose orbital parameters evolve through interactions with nearby PBHs—either isolated singles or loose binaries within these halos and (iii) unperturbed PBH binaries residing outside halos.
For the monochromatic case, we extend the analysis of Ref.~\cite{Aljaf:2024fru} 
by including both hard and soft binaries in the binary-single interaction channel.
For the capture channel, we use the direct results from that reference, and we also 
perform simulations for unperturbed binaries outside halos. The simulations are carried 
out for PBHs with mass $m_{\rm PBH} = 30 \, M_{\odot}$ inside halos spanning 
$10^{4}$-$10^{15} \, M_{\odot}$, and the results are then rescaled to obtain merger 
rates across the full mass range $5$-$80 \, M_{\odot}$ that is currently constrained 
by LVK observations (see discussion in ~\ref{subsec:rates_monochromatic}). 

For the log-normal PBH mass-distribution we have implemented a similar set of simulations as described in Sec.~\ref{subsec:rates_log-normal}. 
Irrespective of the exact mass assumptions, PBH binaries whose orbital properties evolve via interactions with other PBHs inside halos, can significantly enhance the the total PBH merger rate compared to the merging binaries formed from direct captures.  
Our simulations of the PBH binaries interacting with other near-by PBHs account for both the distribution of the orbital properties of the binaries (semi-major axis and eccentricity distributions) and for the relevant environmental factors (density of objects, relative velocities and interacting masses). In our simulations the dark matter halos' properties within which these interactions take place are fully evolved from redshift of 12 to our present time (see discussion in Sec.~\ref{sec:rates} and Refs.~\cite{Aljaf:2024fru, Aljaf:2025dta}). 
We present the total merger rates of PBHs from either a monochromatic or a log-normal mass distribution at redshift of zero in Fig.~\ref{fig:Total_rate_vs_mPBH}, for different mass-choices. Given that these rates represent the most minimal PBH merger rates taking place inside dark matter halos our limits are robust and conservative.

We derive our limits by analyzing the VLK binary black hole primary mass $m_{1}$ and mass-ratio $q$ distributions. We focus on the 72 merger events detected by LVK during the observing runs O1-O3 with a secondary mass $m_{2} \geq  4 \, M_{\odot}$ and with a false alarm rate of FAR $ < 1 \, \textrm{yr}^{-1}$. Following Ref.~\cite{Bouhaddouti:2024ena}, we can use histograms of the $m_{1}$- and $q$-distributions, as those shown in Figs.~\ref{fig:m1_and_q_histrohram_35Msun_monochromatic},~\ref{fig:m_1_and_q_double_best_fit_PBH_normalization},~\ref{fig:m_1_and_q_triple_best_fit_PBH_normalization} and~\ref{fig:m_1_and_q_log-normal_PBH_best_fit_normalization}.
The observed black hole binaries are modeled to be the sum of two distinct populations of binaries.
Most black hole binaries are assumed to have their $m_{1}$, $q$ and comoving rate density distribution properties described by simple power laws. We refer to that population as PL population of binaries (see discussion in Sec.~\ref{sec:BBHs_limits} also Refs.~\cite{PhysRevX.13.011048, Bouhaddouti:2024ena}).
The second population is merging PBH binaries.
Such an analysis of the LVK observations allows for a rapid search of the range potentially describing the underlying binary black hole population.  

Our limits presented in Figs.~\ref{fig:PBH_lims} and~\ref{fig:chi_sqr_map} and in Table~\ref{table:chi_sqr_PP_results}, constrain the dark matter abundance in PBHs to a fraction by mass of $10^{-4} < f_{\textrm{PBH}} < 10^{-2}$. 
These limits derived directly from the GW observations, probe directly the existence of PBHs in the late stages of the universe and thus provide one of the most robust and direct ways to search for them.
In addition, we get some of the tightest limits in the literature on the abundance of PBHS with mass in the range of 3-40 $M_{\odot}$ (see also \cite{Carr:2023tpt}).

As the dominant contribution to the PBH merger rate comes from the 
of existing PBH binaries when interacting with other PBHs, our limits depend on the assumed fraction of 
PBHs in binaries $f_{\rm{PBH \, binaries}}$, at their creation in the early universe (before the formation of dark matter halos). For this reason we allow for a range of options on that fraction; showing in Fig.~\ref{fig:fPBH_and_fPBHbinary} how varying that parameter affects our limits (see also discussion in Sec.~\ref{sec:BBHs_limits}).  

With upcoming results from the LVK fourth run and future final design sensitivity run, we expect the total number of black hole binaries to increase by nearly a factor of ten. 
That improvement alone, will allow us to derive limits on the 
abundance of PBHs that is going to be at least a factor of three tighter. 
That is if the second peak in the LVK data (around masses of $m_{1} = 35 \, M_{\odot}$), is simply a selection effect due to the LVK interferometers' sensitivity and not a true feature of the underlying black hole binaries' population.
Even with the current very low statistics of the LVK observations, we find a minor statistical preference for a population of PBHs with masses $\sim 30 \, M_{\odot}$ 
(for a monochromatic case) for $f_{\textrm{PBH}} \simeq 10^{-2}$ (see Fig.~\ref{fig:chi_sqr_map}). 
Moreover, the future observations by LVK will allow us to probe for black holes with masses $< 4 M_{\odot}$, i.e. in the low-mass gap, a parameter range conventionally though devoid of regular stellar origin black holes. 
With future observations one may also be able to include information on the effective spins of the detected binaries. We leave such a possibility for future work. 

We have made publicly available our results for the PBH merger rates and for the derived PBH limits through \texttt{Zenodo} in  \url{https://zenodo.org/records/14768441}. Additionally, the LVK $m_{1}$- and $q$-histograms used to derive the PBH limits in this work are in \url{https://zenodo.org/records/14675445} (associated to Ref.~\cite{Bouhaddouti:2024ena}). 

\begin{acknowledgments}
The authors would like to thank David Garfinkle and Iason Krommydas for valuable discussions. 
MEB, MA and IC are supported by the National Science Foundation, under grant PHY-2207912.

\end{acknowledgments}

\appendix
\section{Halos properties}\label{app:halo_properries}
The PBH density at particular  redshift follows the NFW dark matter profile, which reads as,
\begin{equation}
\rho_{\textrm{NFW}}(r) = \frac{\rho_s}{\left(r / R_s \right) \left(1 + r / R_s\right)^2}.
\end{equation}

Integrating this density over the volume of a sphere with virial radius $R_{\textrm{vir}}$ gives us the enclosed mass of the halo,
\begin{equation}
M(R_{\textrm{vir}}(z)) = 4 \pi R_s^3 \rho_s\, g(C(z)),
\end{equation}
where $C(z)$ is the concentration parameter of the halo as a function of redshift and is defined as $C(z)=R_{\textrm{vir}}(z) / R_s(z)$.

The $g(C(z)$ and and $f(C)$ appearing in Eq.(\ref{captures_eq}) are functions of the parameters are defined respectively as,
\begin{equation}
\begin{aligned}
    g(C(z)) &= \ln\left(1 + C(z)\right) - \frac{C(z)}{1 + C(z)} \textrm{ and} \\
    f(C(z)) &= 1 - \frac{1}{\left(1 + C(z)\right)^3}.
\end{aligned}
 \end{equation}

The halo mass function quantifies the number of dark matter halos per unit mass per comoving volume of the Universe which is defined as, 
\begin{equation}
\frac{d n}{d \ln M}=M \cdot \frac{\rho_{m_0}}{M^2} f(x)\left|\frac{d \ln x}{d \ln M}\right|. 
\end{equation}
$n$ is the number density of dark matter halos, $M$ is the halo mass, $\rho_{m_0}$ is the mean density of matter, and $f(x)$ is a function related to the geometrical conditions for the overdensities at the collapse time of the halo.  In this paper Press-Schechter \cite{Press:1973iz} halo mass function is used for $f(x)$ which reads as,
\begin{equation}
f(x)=\sqrt{\frac{2}{\pi}} \frac{\delta_c}{x} \exp \left(-\frac{\delta_c^2}{2 x^2}\right),
\end{equation}
where $\delta_c= 1.686$ is the critical overdensity for spherical collapse.

Finally, the  eccentricity functions appearing in the orbital evolutions equations (Eq. \ref{eq:evol_a} and \ref{eq:evol_e}) of the binary-single interactions  are defined as, 
\begin{equation}
F(e)=\left(1-e^2\right)^{-7 / 2} \cdot\left(1+\frac{73}{24} e^2+\frac{37}{96} e^4\right),
\end{equation}
and
\begin{equation}
D(e)=\left(1-e^2\right)^{-5 / 2} \cdot\left(e+\frac{121}{304} e^3\right) .
\end{equation}


\section{Hard PBH binaries inside dark matter halos}
\label{app:Hard_Binaries_fractions}

As described in the main text, only the hard PBH binaries with $a \leq a_{h}$ will harden (become dynamically tighter) via binary-single interactions and can merge within a Hubble time.
The fraction of all PBH binaries that are hard depends on their environments, as in Eq.~(\ref{eq:a_h}), $a_{h}$ depends on the dispersion velocity $v_{\textrm{disp}}(r, t)$ of surrounding PBH objects. 
In this appendix we show results on how exactly the various dark matter halo environments affect the fraction of hard PBH binaries $f_{\textrm{hard}}$. 

In massive dark matter halos, the dispersion velocity $v_{\textrm{disp}}(r, t)$ is high. This means that there is a smaller fraction of hard binaries $f_{\textrm{hard}}$, compared to smaller halos, where the  velocity dispersions are lower. 
In fact in the smallest halos that we study, all PBH binaries are hard. 
Furthermore, the fraction of hard binaries in an evolving halo also changes with time. 
As the dark matter halos grow in mass, early in the halo's evolution, there is a larger fraction of hard binaries, leading to higher merger rates. 
This evolution of $f_{\textrm{hard}}$ is most prominent in halos that grow to be massive by $z=0$. 
In the halos that have low mass at $z=0$, almost all binaries are hard, throughout the simulation time. 

Different choices on the value of the monochromatic PBH mass, $m_{\textrm{PBH}}$, affect our results as well.  A larger value of $m_{\textrm{PBH}}$, leads to a larger value on $a_h$, which in turn increases the fraction of hard binaries and their merger rate due to binary single interactions.  
We remind the reader that the two-body capture rate does not depend on the PBH mass. However, it does depend on $v_{\textrm{disp}}$, with increasing rates at high redshifts. A full analysis of these effects has been recently done in Ref.~\cite{Aljaf:2024fru}.

In Table~\ref{tab:fhard}, we provide the derived 
$f_{\textrm{hard}}$ at $z=0$, for a range of halo masses (first column). For the most massive halos there is a gradient on the value of $v_{\textrm{disp}}(r, t)$ as one moves away from the center of the halo, which affects the value of $a_{h}$ (third column) and $f_{\textrm{hard}}$ (fourth column).  

    \begin{table}
    \centering
    \begin{tabular}{|c|c|c|c|}
        \hline
        Halo Mass ($M_{\odot}$) & Shell Index & $a_{h}$($au$) & $f_{\textrm{hard}}$ (\%)  \\
        \hline
        $ 10^{3}$ & - & $4.48 \times 10^{5}$ & 100. \\
                \hline
        $ 10^{6}$ & 1 & $2.08 \times 10^{-1}$ & 99.6 \\
                \hline
        $ 10^{6}$ & 3 & $1.79 \times 10^{-2}$ & 41 \\
                \hline
        $ 10^{6}$ & 5 & $3.17 \times 10^{-2}$ & 61 \\
                \hline
        $ 10^{9}$ & 1 & $7.07 \times 10^{-2}$ & 87 \\
                \hline
        $ 10^{9}$ & 6 & $4.20 \times 10^{-4}$ & 0.67 \\
                \hline
        $ 10^{9}$ & 10 & $3.17 \times 10^{-4}$ & 0.46 \\
                \hline
        $ 10^{12}$ & 1 & $1.07 \times 10^{-2}$ & 27 \\
                \hline
        $ 10^{12}$ & 6 & $1.79 \times 10^{-5}$ & 0.022 \\
                \hline
        $ 10^{12}$ & 10 & $3.17 \times 10^{-6}$ & 0.003 \\
                \hline
        $ 10^{15}$ & 1 & $2.26 \times 10^{-3}$ & 5.1 \\
                \hline
        $ 10^{15}$ & 6 & $1.21 \times 10^{-6}$ & 0 \\
                \hline
        $ 10^{15}$ & 10 & $3.17 \times 10^{-8}$ & 0 \\
                \hline
        \end{tabular}
        \caption{For a range of dark matter halo masses (first column) and at different radial distances (second column) from the center of the halos, we provide in the last column the fraction of hard binaries, $f_{\textrm{hard}}$ (in \%). We give results at $z=0$. We sampled N$=10^{5}$ binaries in each halo. We used the case of a monochromatic $30 \, M_{\odot}$ PBHs. For reference we also provide the value of $a_{h}$ for the given environment (second to last column).}     
        \label{tab:fhard}
        \end{table}
In these calculations, for the dark matter halos with mass of $10^3 \, M_\odot$ we assumed a uniform density evaluated from the NFW profile at $R_{\textrm{vir}}/2$. For the smallest halos there is no radial gradient on the velocity dispersion, as a PBH's crossing time through the halo is too small compared to the evolution timescale. For $10^6 \, M_\odot$ halos we used five radial shells, while for the more massive $10^{12} \, M_\odot$ and $10^{15} \, M_\odot$ halos, ten radial shells. The shell index starts at one for the innermost shell and increases outward. For more details on the structure of the simulations see Ref.~\cite{Aljaf:2024fru}.

\section{The orbital properties of PBH binaries before halo formation}
\label{app:BBHs_Initial_Semimajor_Distribution}
Since our goal is to simulate the evolution and merger rates of PBH binaries, we need to know the initial value of their orbital properties $(a_0, \, e_0)$. In this appendix, we provide the initial distribution of $(a_0,\, e_0)$, of PBH binaries at their formation. 
For the case of PBH binaries with monochromatic mass, a detailed discussion about the distribution of their orbital parameters is given in \cite{Ali-Haimoud:2017rtz,Kavanagh:2018ggo,Liu:2018ess,Franciolini:2021xbq}. 
We follow the exact assumptions outlined in \cite{Franciolini:2021xbq} (see also Ref.~\cite{Aljaf:2024fru}), to calculate the the initial distribution of $(a_0, \, e_0)$ of PBH binaries. 
In this work, we also simulate the merger rates of PBH binaries following a log-normal mass distribution. For such binaries the distribution of their orbital properties depends on their mass distribution at formation. 
Thus, we need to modify the formulation and assumptions provided in \cite{Franciolini:2021xbq} and extend it to the case of a log-normal distribution. We simply take expectation values for terms that include the PBH's mass.  

The mean separation between an initial pair of PBHs is, 
\begin{equation}\label{xbar}
\bar{x} =\left(\frac{3 (m_{\mathrm{PBH,\, 1}}+m_{\mathrm{PBH,\, 2}})}{8 \pi f_{\mathrm{PBH}} \,\rho_{\mathrm{eq}}}\right)^{1 / 3},
\end{equation} 
where $\rho_{\mathrm {eq}}$ is the average energy density in the matter-radiation equality (at $z\simeq 3450$) and $m_{\mathrm{PBH,\, 1}}$ and $m_{\mathrm{PBH, \, 2}}$ are the masses of the two PBHs forming the binary. 

The distribution of the semi-major axis depends on $\bar{x}$ and the rescaled angular momentum $j \equiv \sqrt{1-e^2}$ that read as follows,
\begin{eqnarray}
 P(j) &=&\frac{y(j)^2}{j\left(1+y(j)^2\right)^{3 / 2}} \; \; \; \;  \textrm{and} \\
y(j) & = & \frac{j}{0.5\left(1+\sigma_{\mathrm{eq}}^2 / f_{\mathrm{PBH}}^2\right)^{1 / 2}(x / \bar{x})^3}.
\label{eq:P_j}
\end{eqnarray}
The parameter $\sigma_{\mathrm{eq}} \approx 0.005$, is  the variance of the Gaussian large-scale density fluctuations during matter-radiation equality. 
This distribution is the sum of the effect by torques from nearby PBHs and matter perturbations during the PBH binary's formation \cite{Liu:2018ess}. Ultimately, the distribution that characterizes both $j$ and the semi-major axis $a$ can be expressed as,
\begin{equation}\label{P_a_j}
P(a, j)=\frac{3  \cdot a^{-1 / 4}}{4}\left(\frac{f_{\mathrm{PBH}}}{\zeta \bar{x}}\right)^{3 / 4} P(j) \exp \left[-\left(\frac{x(a)}{\bar{x}}\right)^3\right],
\end{equation}
with
\begin{equation} \label{xa}
x(a) =\left(\frac{3 a ( m_{\mathrm{PBH, \, 1}}+ m_{\mathrm{PBH, \, 2}})}{8 \pi \zeta \,\rho_{\mathrm{eq}}}\right)^ {1 / 4}
\end{equation}
and  $\zeta=0.1$ \cite{Ali-Haimoud:2017rtz}.

The parameters $\bar{x}$ and $x(a)$, can therefore be calculated for any given pair of ($m_{\mathrm{PBH,\, 1}}, \, m_{\mathrm{PBH, \, 2}}$). However, as described in the main text, when PBHs follow a log-normal mass-distribution with the function $\psi(m)$ of Eq.~(\ref{eq:MassPDF_lognormal}), the values for ($m_{\mathrm{PBH, \, 1}}, \, m_{\mathrm{PBH, \, 2}}$) span a range. 
To account for such a range, we evaluate $\bar{x}$ and $x(a)$ analytically by integrating over the respective mass distribution functions $\psi(m_{1})$ and $\psi(m_{2})$ as, 
\begin{eqnarray}
\left\langle\bar{x}\right\rangle&=&\int_{m_{1_{min}}}^{m_{1_{max}}} dm_{1} \int_{m_{2_{min}}}^{m_{2_{max}}} dm_{2}
\, \psi(m_{1}) \, \psi(m_{2}) \nonumber \\
&\times& \left(\frac{3 (m_{1}+m_{2})}{8 \pi f_{\mathrm{PBH}} \, \rho_{\mathrm{eq}}}\right)^{1 / 3}.
\label{eq:xbar_av}
\end{eqnarray}
and
\begin{eqnarray}
   \left\langle x(a)\right\rangle &=&\int_{m_{1_{min}}}^{m_{1_{max}}} dm_{1} \int_{m_{2_{min}}}^{m_{2_{max}}} dm_{2}  \, \psi(m_{1}) \, \psi(m_{2}) \nonumber \\
   &\times& \left(\frac{3 a ( m_{1}+ m_{2})}{8 \pi \zeta \rho_{\mathrm{eq}}}\right)^ {1 / 4}.
   \label{eq:xa_av}
\end{eqnarray}
With, $m_{1}$ and $m_{2}$ the $m_{\mathrm{PBH, \, 1}}$ and $m_{\mathrm{PBH,\, 2}}$ respectively. 

\begin{figure}[h!]
    \centering
\includegraphics[width=1\linewidth]{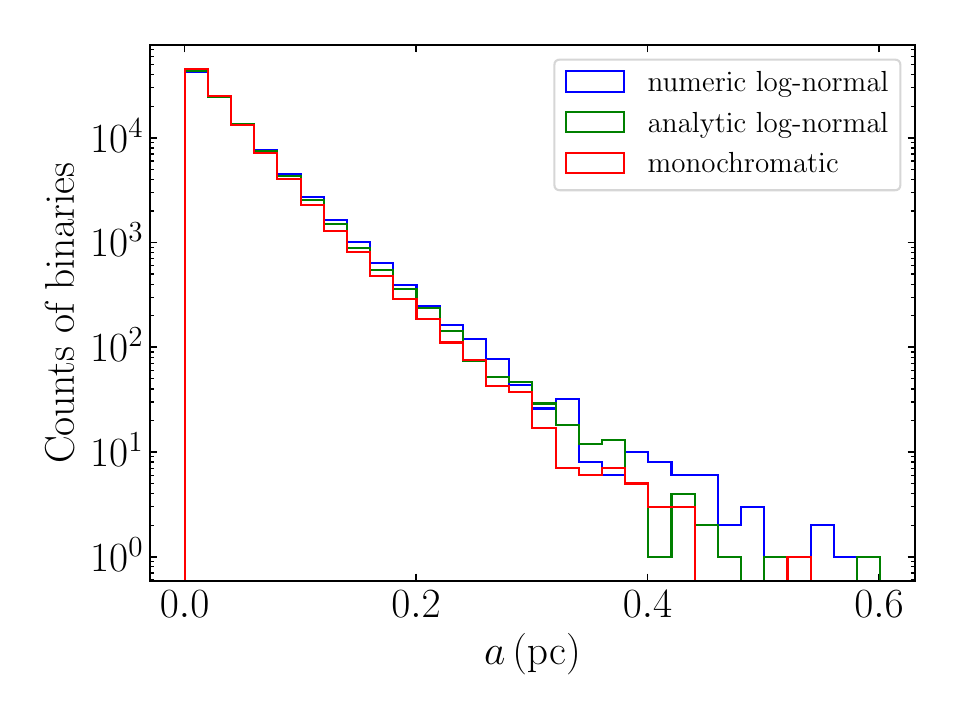}
    \caption{The initial semi-major axis of PBHs, calculated for monochromatic PBHs with $m_{\textrm{PBH}}=30 M_\odot$ (red line) and for a log-normal distribution with $\mu =30)$ and $\sigma=0.6$. For the log-normal distribution we present results through two different prescriptions; a fully numerical Monte Carlo simulation ``numeric'' (shown by the blue line) and an analytical prescription (green line). See text for more details.}
    \label{fig:PBBHs_a}
\end{figure}

In Fig.~\ref{fig:PBBHs_a}, we show in the green-lined histogram the semi-major axis distribution for the PBH binaries at their initial conditions. We assumed a log-normal mass distribution with $\mu =\textrm{ln}(30)$ and $\sigma =0.6$. We also show for comparison, in the red-lined histogram the equivalent semi-major axis distribution for binaries made out of monochromatic $m_{\textrm{PBH}} = 30 M_{\odot}$ PBHs. We simulated $10^{5}$ PBH binaries. 
 
While Eqs.~(\ref{eq:xbar_av}) and~(\ref{eq:xa_av}), provide an analytical prescription to get the PBH binaries' semi-major axis distribution, we have also run a Monte Carlo simulation of these binaries' distribution (shown by the blue-line ``numeric log-normal'' histogram). 
For the numeric-log-normal prescription, we generated the masses (not taking the mean of the mass terms),  $m_{\textrm{PBH, 1}}$ and $m_{\textrm{PBH, 2}}$ in the expression for $\bar{x}$ and $x(a)$ given by Eqs.~(\ref{xbar}) and~(\ref{xa}). We applied the inverse transform sampling method to the log-normal distribution. 
This calculation lead to the generation of unique values for $\bar{x}$ and $x(a)$ (though very similar values) for each binary we sampled. 
This means that, unlike the analytical prescription, we calculated the separation between each binary we sampled and did not assume one mean separation $\bar{x}$ for all binaries as in the case of the analytical log-normal.
The difference between the numerical and the analytical prescriptions on the resulting distributions of the semi-major axes of the PBH binaries at creation, are minute. 

\section{Examples of $m_{1}$ and $q$ histograms from our fits to the LVK data}
\label{app:m1_and_q_histograms}

In Fig.~\ref{fig:m1_and_q_histrohram_35Msun_monochromatic}, we presented the fitted $m_{1}$- and $q$-histograms associated to a combination of a PL and PBH component (using $m_{\textrm{PBH}} = 35 \, M_{\odot}$ as an example). 
While adding a PBH component improves the fit to the $m_{1}$-histogram alone, for masses around its second peak, we noticed that the quality of the fit to the $q$-histogram is always poorer compared to a PL component only. 
For the example of Fig.~\ref{fig:m1_and_q_histrohram_35Msun_monochromatic}, its bottom panel, we get a $\chi^2$ fit to 
the LVK $q$-histogram of $\chi^2_{\textrm{PL 
\& PBH}, \, q} = 19.7$. By comparison fitting the $q$-histogram by a simple PL component we get $\chi^2_{\textrm{PL}, \, q} = 17.8$.

\begin{figure}[ht!]
    \centering
    \includegraphics[width=1.018\linewidth]{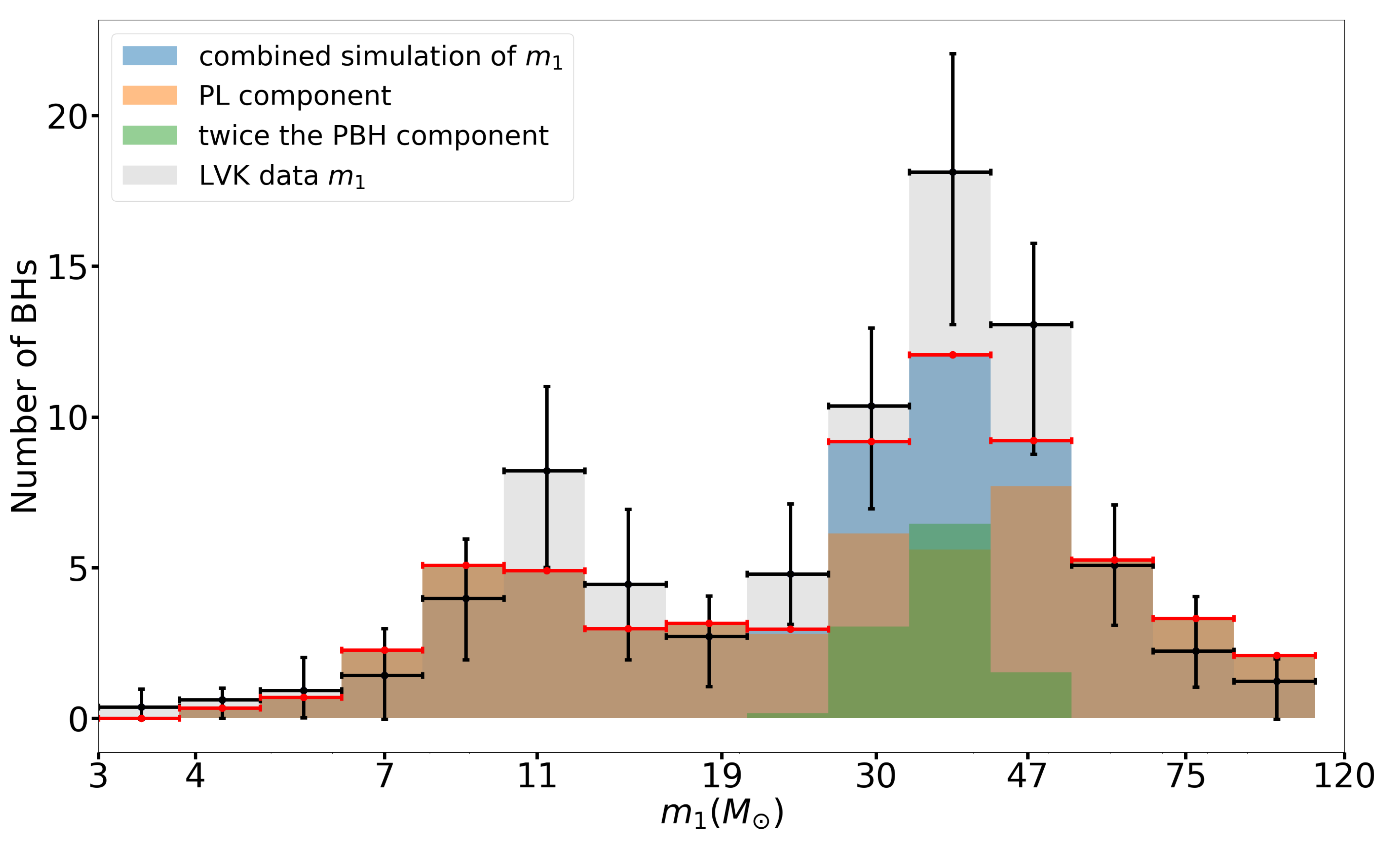}\\
    \vspace{0.1in}
    \includegraphics[width=1\linewidth]{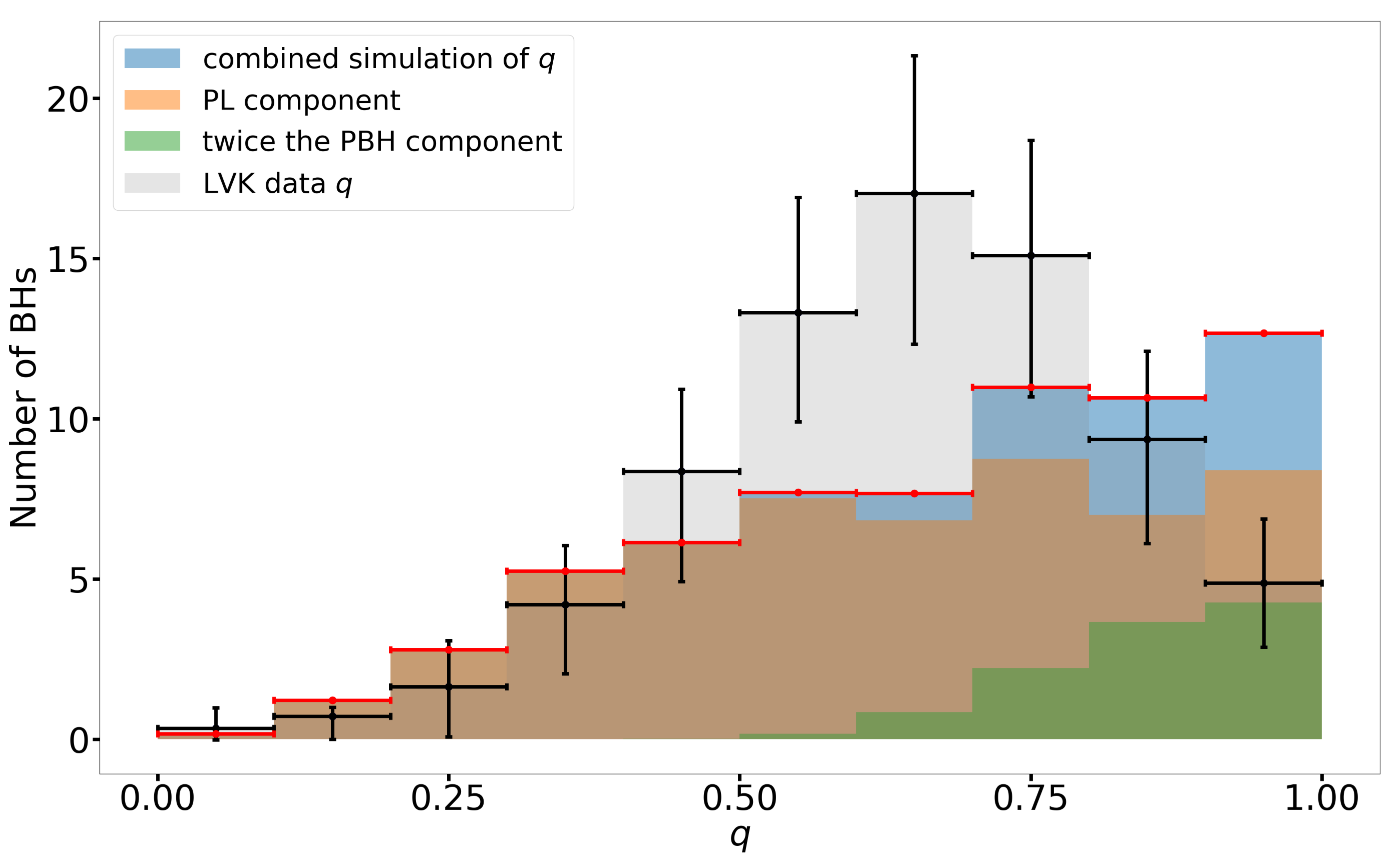}
    \caption{As in Fig.~\ref{fig:m1_and_q_histrohram_35Msun_monochromatic}, we show the normalized $m_1$-histogram (top) and $q$-histogram (bottom) of the simulated (in blue)
and detected by LVK (in gray) black hole binaries with a
SNR> 8. The simulated binaries are a combination of a dominant PL population (in brown) and double the best fit population of PBH binaries of Fig.~\ref{fig:m1_and_q_histrohram_35Msun_monochromatic} (in green). See text for further details.}
    \label{fig:m_1_and_q_double_best_fit_PBH_normalization}
\end{figure}

\begin{figure}[ht!]
    \centering
    \includegraphics[width=1.018\linewidth]{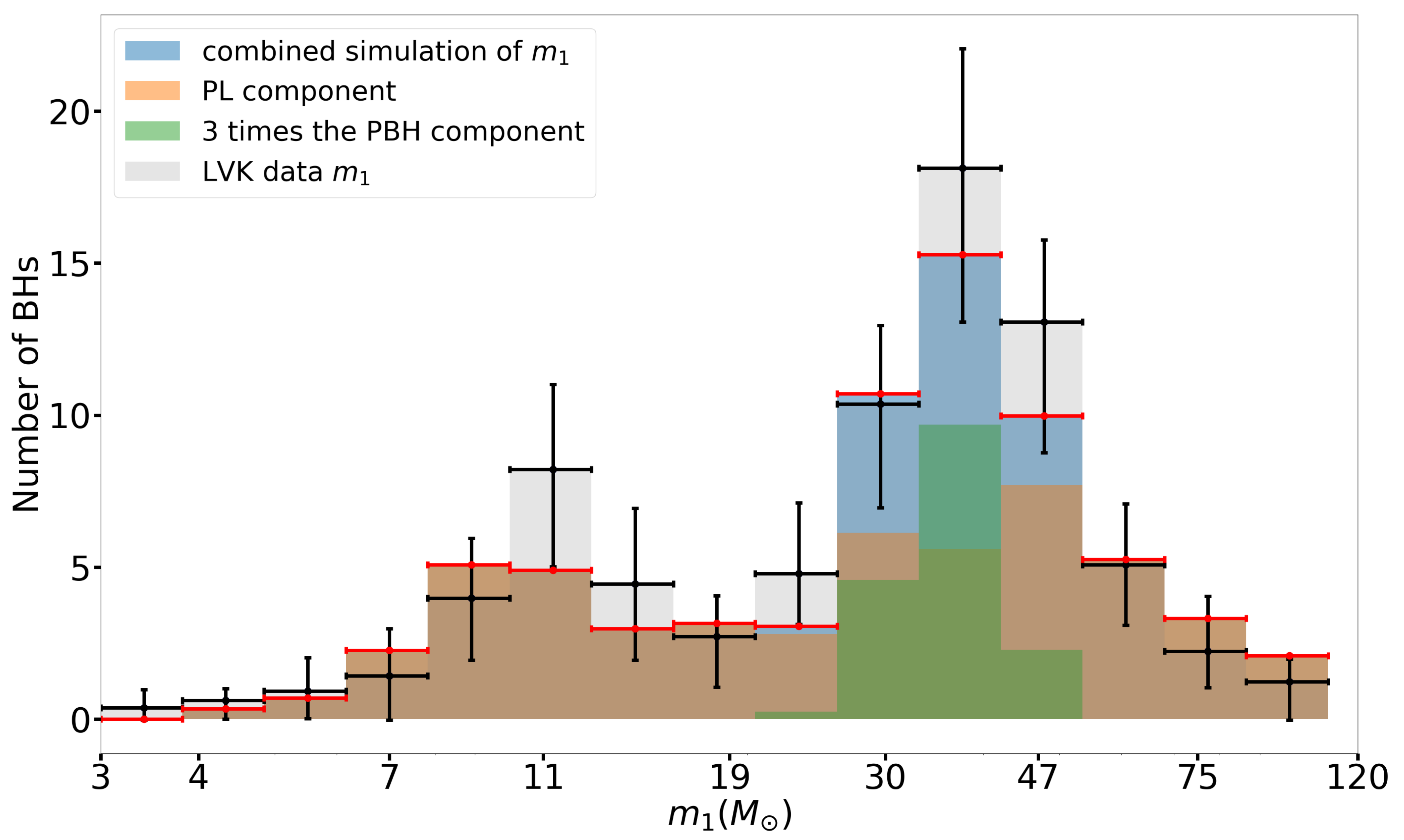}\\
    \includegraphics[width=1\linewidth]{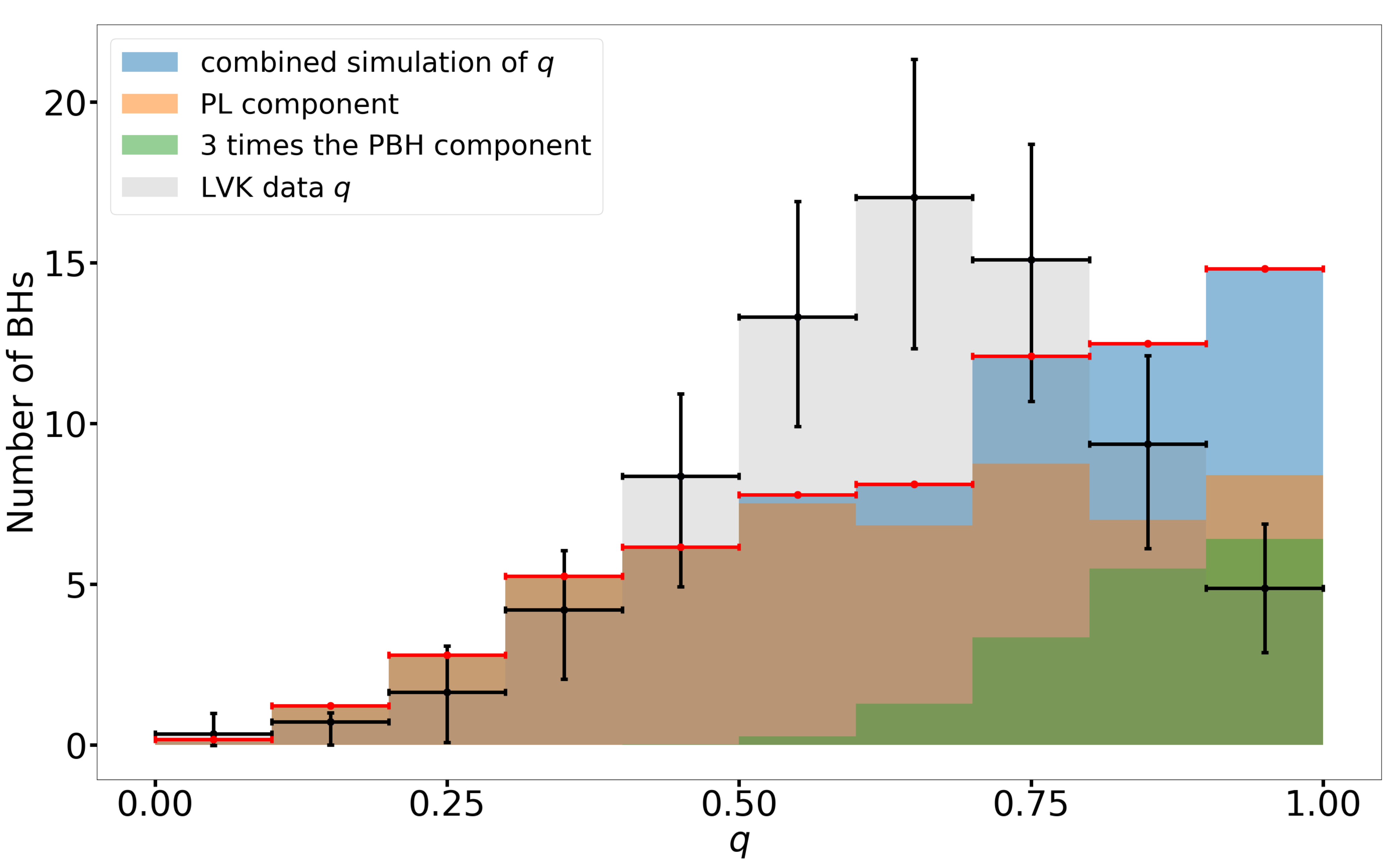}
    \caption{As in Fig.~\ref{fig:m_1_and_q_double_best_fit_PBH_normalization}, but instead using three times the (best fit) normalization for the PBH component that is shown in Fig.~\ref{fig:m1_and_q_histrohram_35Msun_monochromatic}.}
    \label{fig:m_1_and_q_triple_best_fit_PBH_normalization}
\end{figure}

A larger PBH component would lead to a better fit of the $m_1$-histogram. However, the detrimental effect of the PBH component on the fitting of $q$, limits its normalization significantly. 
In Fig.~\ref{fig:m_1_and_q_double_best_fit_PBH_normalization} and Fig.~\ref{fig:m_1_and_q_triple_best_fit_PBH_normalization} we present the $m_1$- and $q$-histograms, using respectively double and triple the best-fit normalization to the PBH component of Fig.~\ref{fig:m1_and_q_histrohram_35Msun_monochromatic}. 
The best-fit for a PL $\&$ PBH component gave 
$\chi^2_{\textrm{PL 
\& PBH best-fit}, \, m_{1}} = 12.2$ and
$\chi^2_{\textrm{PL 
\& PBH best-fit}, \, q} = 19.7$, for $m_{\textrm{PBH}} = 35 \, M_{\odot}$ and $f_{\textrm{PBH}} = 1.1 \times 10^{-3}$.
For the same $m_{\textrm{PBH}} = 35 \, M_{\odot}$, doubling $f_{\textrm{PBH}}$ to $2.2 \times 10^{-3}$ we get, $\chi^2_{\textrm{PL 
\& \,} 2\times\textrm{PBH}, \, m_{1}} = 8.5$ and $\chi^2_{\textrm{PL 
\& \,} 2\times\textrm{PBH}, \, q} = 25.87$.
If we triple the $f_{\textrm{PBH}}$ to $ 3.3 \times 10^{-3}$ we get and $\chi^2_{\textrm{PL 
\& \,} 2\times\textrm{PBH}, \, m_{1}} = 6.43$ and $\chi^2_{\textrm{PL 
\& \,} 2\times\textrm{PBH}, \, q} = 35.27$. This shows that how a poor fit of the $q$-histogram of the LVK observations makes our limits on $f_{\textrm{PBH}}$ significantly more restrictive.

\begin{figure}
    \centering
    \includegraphics[width=1.018\linewidth]{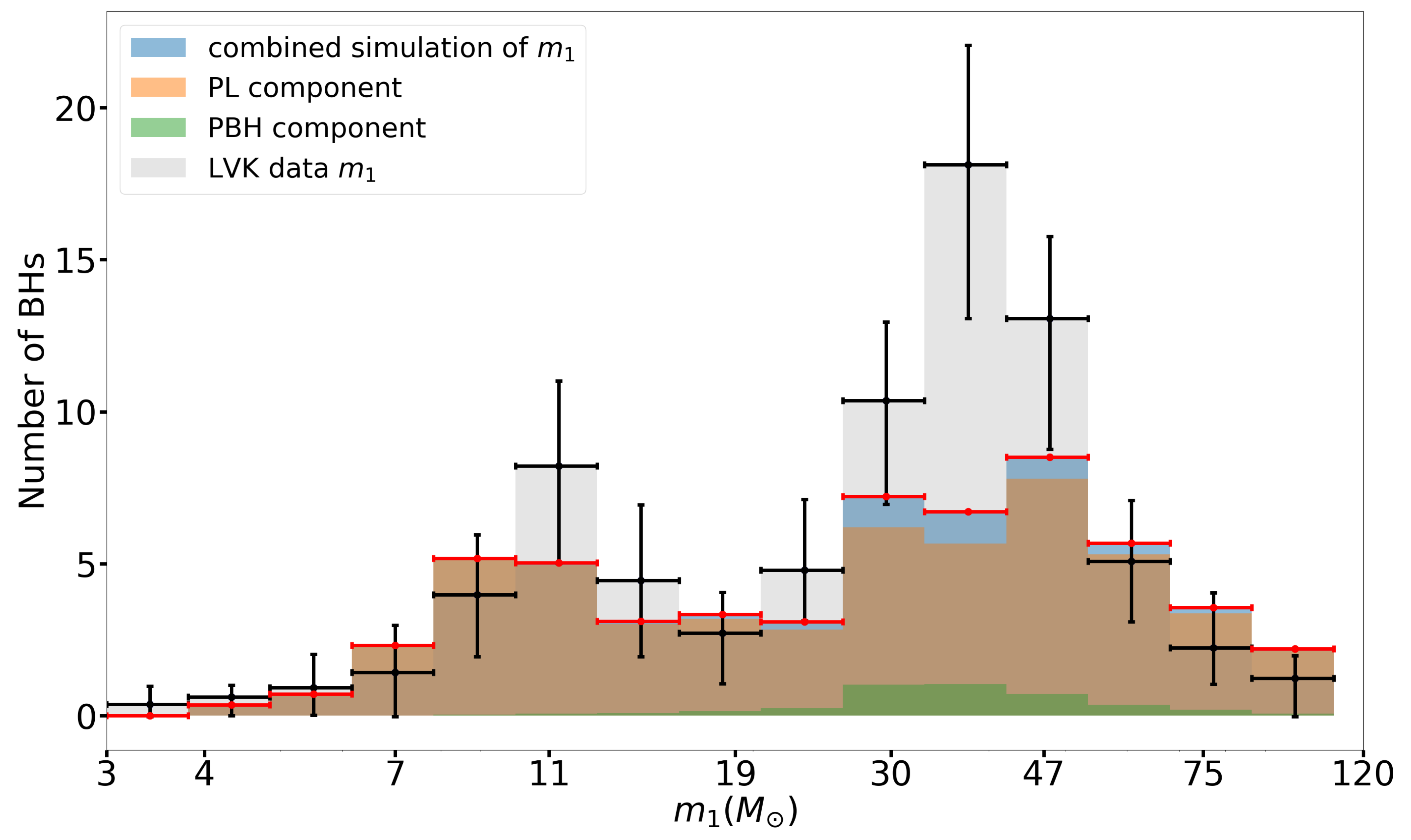}\\
    \includegraphics[width=1\linewidth]{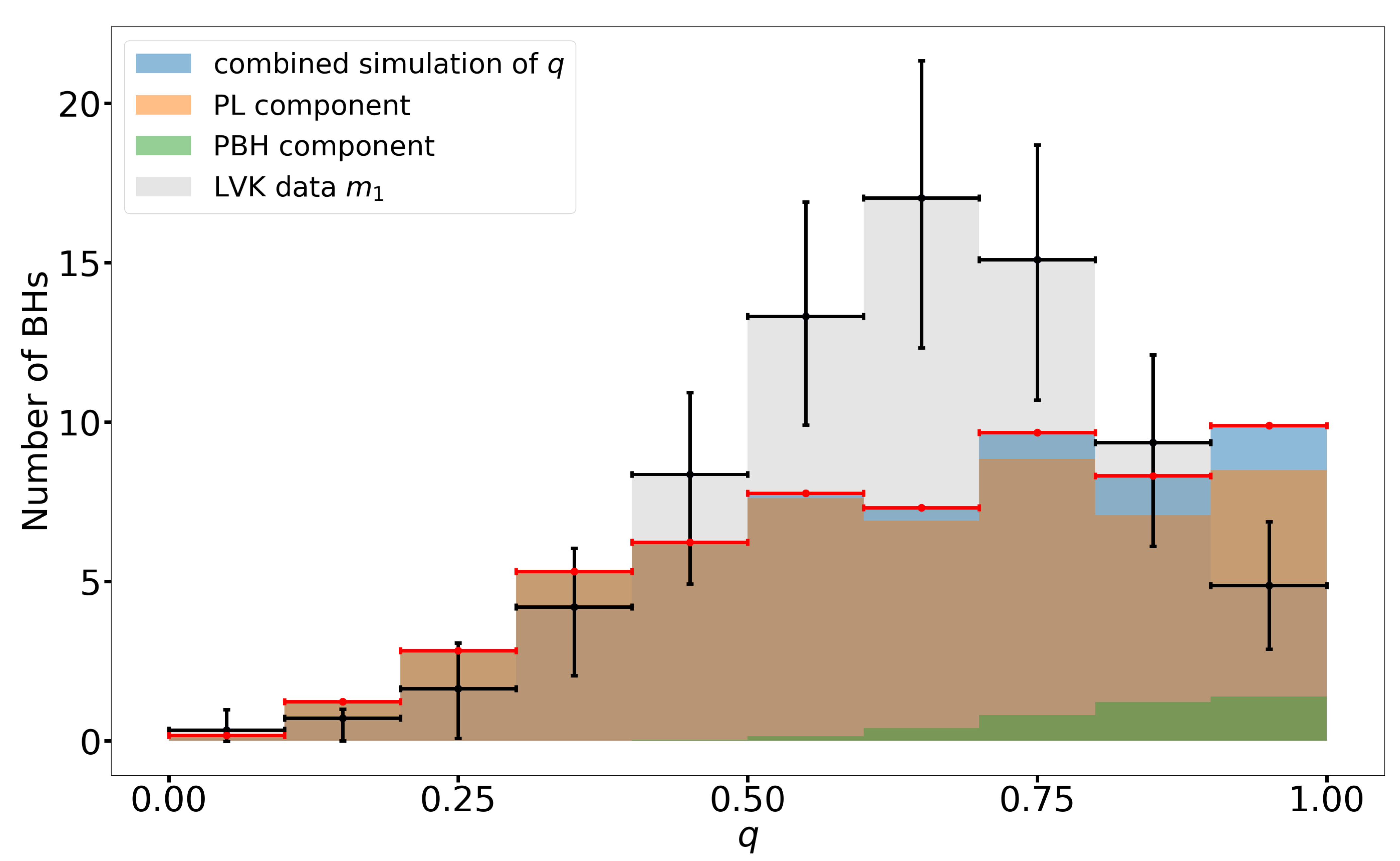}
    \caption{As in Fig.~\ref{fig:m1_and_q_histrohram_35Msun_monochromatic}, we show the normalized $m_1$-histogram (top) and $q$-histogram (bottom) of the simulated (in blue)
    and detected by LVK (in gray) black hole binaries with a
    SNR> 8. The simulated binaries are a combination of a dominant PL population (in brown) and the best fit population generated from a log-normal distribution of PBHs }
    \label{fig:m_1_and_q_log-normal_PBH_best_fit_normalization}
\end{figure}

Finally, in Fig.~\ref{fig:m_1_and_q_log-normal_PBH_best_fit_normalization}, we show the best fit assumptions for a PL and PBH population, where the PBHs follow a log-normal distribution with $\mu =\textrm{ln}(15)$ and $\sigma = 0.6$. 
The best fit normalization is achieved for $f_{\textrm{PBH}} = 1.1 \times 10^{-3}$ (with $f_{\mathrm{PBH \, binaries}} = 0.5$).

\bibliography{main.bib}
\end{document}